\documentclass{jfm}
\usepackage{graphicx}
\usepackage{amsmath}
\usepackage{multirow}
\usepackage{placeins}
\usepackage{todonotes}
\usepackage{textgreek}
\usepackage{mathrsfs}
\usepackage{bm}
\usepackage{enumitem}

\usepackage[normalem]{ulem}

\newcommand\T{\rule{0pt}{2.6ex}}       
\newcommand\B{\rule[-1.2ex]{0pt}{0pt}} 

\newcommand\bu{\boldsymbol{u}}
\newcommand\bU{\boldsymbol{U}}
\newcommand\Ro{\mbox{\textit{Ro}}} 
\newcommand\Ri{\mbox{\textit{Ri}}} 
\newcommand\Bu{\mbox{\textit{Bu}}} 

\def\ee{{\rm e}}
\def\ii{{\rm i}}
\DeclareMathOperator\erf{erf}

\usepackage{bigstrut}
\usepackage{booktabs, multirow} 
\usepackage{soul}
\usepackage{changepage,threeparttable} 
\usepackage{tabularx,ragged2e,booktabs}

\shorttitle{Energy exchanges between 2D front and internal waves}
\shortauthor{S. Kar \& R. Barkan}

\title{{Energy exchanges between a two-dimensional front and internal wave modes}}
\author{Subhajit Kar\aff{1}
\corresp{\email{subhajitkar@mail.tau.ac.il}},
 \and Roy Barkan\aff{1,2}}

\affiliation{\aff{1}Porter School of the Environment and Earth Sciences, Tel Aviv University, Ramat Aviv,  Israel 6997801.
\aff{2}Department of Atmospheric and Oceanic Sciences, University of California, Los Angeles, CA, USA.
 }

\begin{document}

\maketitle

\begin{abstract}
Fronts and near-inertial waves are energetic motions in the upper ocean that can interact and provide a route for kinetic energy (KE) dissipation of balanced oceanic flows. A quasilinear model is developed to study the KE exchanges between a two-dimensional geostrophically-balanced front undergoing strain-induced semigeostrophic frontogenesis and internal wave (IW) vertical modes. The quasilinear model is solved numerically for variable imposed strain magnitudes, initial IW vertical modes, and for both minimum frequency (near-inertial, NI) and high-frequency IWs. 
The front-IW KE exchanges are quantified separately during two frontogenetic stages: an exponential sharpening stage that is characterized by a low Rossby number and is driven by the imposed geostrophic strain, followed by a superexponential sharpening stage that is characterized by an $\mathcal{O}(1)$ Rossby number and is driven by the convergence of the ageostrophic secondary circulation. It is demonstrated that high-frequency IWs quickly escape the frontal zone and are very efficient at extracting KE from the imposed geostrophic strain field through the deformation shear production (DSP) mechanism. Part of the extracted KE is then converted to wave potential energy. On the contrary, minimum frequency IWs remain locked to the frontal zone and therefore exchange energy with the ageostrophic frontal circulation. During the exponential stage, the front-NIW interactions are consistent with those reported in Thomas (\textit{J. Fluid Mech.}, vol. $711, 2012$, pp. $620-640$), where IWs extract KE from the geostrophic strain through DSP and transfer it to the the frontal secondary circulation via the ageostrophic shear production (AGSP) mechanism. During the superexponential stage a newly identified mechanism, `convergence production' (CP), which is directly linked to the convergent secondary circulation, plays an important role in the NIW KE budget. The CP transfers KE from the convergent ageostrophic secondary circulation to the IW, and largely cancels out the KE loss due to the AGSP.
\end{abstract}
\section{Introduction}
Mesoscale geostrophic eddies comprise the largest reservoir of kinetic energy (KE) in the ocean \citep{ferrari2009ocean}. Because their dynamics are constrained by geostrophic and hydrostatic force balanced, they are expected, according to geostrophic turbulence theory \citep{salmon1980baroclinic}, to transfer their KE to larger scales (inverse cascade). The mechanisms that halt that inverse KE cascade, and permit a forward KE cascade to dissipative scale, have been a topic of much debate in oceanography \citep{MJJ05}. 

We focus here on the mechanism first proposed by \cite{gertz2009near}, whereby storm-forced near-inertial waves (NIWs) can interact with mesoscale geostrophic eddies and drain a considerable fraction of their KE. 
To explain this mechanism, \cite{xie2015generalised} constructed an asymptotic theory based on the generalized Lagrangian-mean {framework} (GLM) to study the interactions between NIWs and balanced quasi-geostrophic (QG) flow. \cite{wagner2016three} arrived at a similar NIW-QG coupled system by using an Eulerian-based multiple time scale approach. In both {theories} the NIW dynamics is governed by the so-called YBJ equation \citep{young1997propagation}\footnote{later refined to the YBJ$^+$ equation \citep{asselin2019improved}}. The essential ingredients in these reduced models are the conservation of the total energy (QG + NIW) and the near-inertial wave action (or wave kinetic energy). \cite{rocha2018stimulated} studied the NIW-QG system in coupled numerical simulations of barotropic (2D) turbulence and NIW vertical modes. They demonstrated that any reduction in the horizontal scales of NIWs must be accompanied by an increase in wave potential energy and a subsequent reduction in the kinetic energy of the balanced flow (a mechanism they referred to as {\it stimulated generation}). It is noteworthy, however, that {\it stimulated generation} is only cleanly identified in the GLM framework where the Lagrangian-mean `balanced' flow contains wave-induced contributions. It remains difficult to evaluate {\it stimulated generation} in Eulerian-based numerical models or in-situ measurements. 

\cite{thomas2020near} and \cite{thomas2021forward} used idealized numerical simulations of Boussinesq flow in the small Rossby number parameter regime, characteristic of QG dynamics, and showed that when the wave amplitude is much larger than that of the QG flow (i.e., strong-wave limit), NIWs facilitate the downscale KE cascade of the balanced flow. Furthermore, \cite{thomas2021forward} demonstrated that when wave and balanced flow amplitudes are comparable, the downscale cascade is reduced and results in the accumulation of KE at large scales. \cite{xie2020downscale} found similar modifications to the KE cascades in numerical simulations of the NIW-QG reduced model. 

Other numerical studies have examined more realistic configurations and investigated the balanced flow evolution under the influence of high-frequency wind forcing. For example, \cite{taylor2016forced} simulated an eddy-permitting wind-driven channel flow and showed that the Reynolds stresses associated with NIWs can provide a route for KE dissipation of mesoscale geostrophic flow. \cite{barkan2017stimulated} used a similar configuration albeit with a much higher spatial resolution that allowed to simulate submesoscale currents, which are characterized by a much larger Rossby number \citep{TTM08, mcwilliams2016submesoscale}. They demonstrated that the internal wave-induced energy pathways include two routes - first, direct energy extraction from the mesoscale flow by the externally forced NIWs followed by an internal wave downscale KE cascade to dissipation, and second, a stimulated imbalance process that involves an IW triggered forward energy cascade from meso to submeso time scales. More recently, using realistically forced ocean simulations, \cite{barkan2021oceanic} demonstrated that the most significant energy exchanges between NIWs and subinertial motions are localized in strongly baroclinic submesoscale fronts and filaments that dynamically depart from geostrophic balance.
 
The effects of strongly baroclinic fronts on the polarization relations of NIWs and the subsequent energy exchanges were studied by \citet[hereinafter T12]{thomas2012effects}. T12 developed an idealized model for an unbounded two-dimensional front and showed that NIWs efficiently extract energy from a geostrophic deformation field and transfer it to the ageostrophic circulation that develops spontaneously during frontogenesis.  \cite{whitt2015resonant} used a slab mixed layer model to illustrate that inertial oscillations can exchange energy periodically with a unidirectional, laterally sheared geostrophic flow, and \cite{jing2017energy} pointed out that it is the geostrophic strain that makes this energy transfer permanent. 

In this study, we extend the work of T12 and examine the energy exchanges between IW vertical modes and a two-dimensional front undergoing strain-induced semigeostrophic frontogenesis \citep[hereinafter HB72]{hoskins1972atmospheric}. 
In our model, the frontal sharpening process occurs in two stages - an \textit{exponential} growth stage, driven by the imposed geostrophic deformation field, followed by a \textit{superexponential} growth stage, driven by the convergent ageostrophic secondary circulation (ASC). This \textit{superexponential} growth stage is characteristic of oceanic submesoscale frontogenesis (e.g. \citealp{barkan2019role}). 
We identify a new mechanism, `Convergence Production' (CP), whereby the convergent ASC during submesoscale frontogenesis allows NIWs to efficiently extract KE from the front.
Nonetheless, the CP mechanism has been previously described in the context of cross-scale energy transfers during frontogenesis \citep{srinivasan2021cascade}, but never before discussed in the context of energy exchanges between NIWs and fronts. 
It is shown that CP is the dominant KE extraction mechanism by all of the NIW modes considered during the \textit{superexponential} growth stage. 



The paper is organized as follows. In \S \ref{prob_config} and \S \ref{sec_wave_ke} we discuss the configuration used to study front-IW interactions, utilizing the mathematical framework developed in \citet[hereinafter ST13]{shakespeare2013generalized}, distinguishing between minimum frequency (near-inertial) and high-frequency IWs. 
{The details of the numerical setup are provided in \S \ref{numerical_model}}.
{In \S \ref{results}, we discuss the evolution of the mean flow and compare the numerical solution with 2D semi-analytical frontogenesis solutions.}
A detailed analysis of the front-IW energy exchanges is shown in \S \ref{sec_energetics}. Finally, in \S \ref{summary}, we summarize our findings and draw connections to realistic ocean scenarios. 

\section{Problem configuration}
\label{prob_config}
An idealized configuration is developed to study front-IW energy exchanges. The configuration consists of a $2$D (i.e., invariant in the $x$-direction) geostrophically-balanced front undergoing strain-induced frontogenesis, to which we add IW vertical modes in a bounded domain of width $L$ and depth $H$.  Our goal is to define `balanced' mean-flow evolution equations that solely describe frontogenesis in a way that is analytically tractable, numerically solvable, and excludes fast internal-wave motions that could be generated due to geostrophic adjustment and/or spontaneous emission. Such mean-flow would evolve on a slower time scale than the IW vertical modes, allowing for an unambiguous quantification of the energy exchanges. The dynamics of the mean-flow and the IW vertical modes are governed by the hydrostatic, Boussinesq equations of motion for a rotating fluid under the $f$-plane approximation.
\subsection{Mean-flow evolution equations}
\label{mean_flow_config}
The mean-flow velocity ($\overline{\bU}\equiv(\overline{U},\overline{V},\overline{W})$), buoyancy ($\overline{B}$), and pressure ($\overline{P}$) fields take the form
\begin{subequations}
\begin{align}
\label{eq1a}
\overline{U} &= \alpha x + {U}(y, z, t), \\
\overline{V} &= -\alpha y + {V}(y, z, t), \\
\overline{W} &= {W}(y, z, t), \\
\overline{P} &= {P}_0(x,y) + {P}(y, z, t), \\
\label{eq1e}
\overline{B} &= {B}(y, z, t),
\end{align}
\end{subequations}
where $\alpha$ denotes a spatially and temporally uniform large-scale geostrophic strain, which is used to initiate frontogenesis, 
and the velocity components $ \overline{U}, \overline{V}$ and $\overline{W}$ are oriented in the $\hat{x}, \hat{y}$ and $\hat{z}$ directions, respectively. 
The mean-flow pressure $\overline{P}$ consists of a pressure field that balances the geostrophic deformation flow ${P}_0 = -\rho_0 
\big[ \alpha^2(x^2+y^2)/2 + f\alpha x y \big]$ and $P$, which is in hydrostatic balance with the mean-flow buoyancy ${B} = -{g \rho^\prime}/{\rho_0}$ ($\rho^\prime$ is the mean-flow density perturbation relative to the reference density $\rho_0$, and $g$ is the gravitational acceleration).

The resulting mean-flow evolution equations are:
\begin{subequations}
\label{gov_eqs}
\begin{align}
\label{eq_gov_a}
\frac{D U}{D t} - f V + \alpha U &= 0, \\
\label{eq_gov_b}
\frac{D V}{D t} + f U - \alpha V &= -\frac{\partial P}{\partial y}, \\
\label{eq_gov_c}
0 &= -\frac{\partial P}{\partial z} + B, \\
\label{eq_gov_d}
\frac{D B}{D t} &= 0, \\
\label{eq_gov_e}
\frac{\partial V}{\partial y} + \frac{\partial W}{\partial z} &= 0,
\end{align}
\end{subequations}
where the material derivative is defined as
\begin{align}
\label{mat_der}
    \frac{D}{Dt} 
    \equiv \frac{\partial}{\partial t} + (V- \alpha y) \frac{\partial}{\partial y} + W \frac{\partial }{\partial z}.
\end{align}
As discussed in ST13, the above equations can be solved for an initially uniform potential vorticity (PV) distribution. The solution contains three processes: 
\smallskip

\begin{enumerate}
\renewcommand{\theenumi}{\roman{enumi}}
\item  Semigeostrophic frontogenesis (i.e., HB72),
\item  Spontaneous internal-wave (IW) emission due to the external geostrophic strain, 
\item  Geostrophic adjustment and IW excitation  (i.e., \citealt{blumen2000inertial}).
\end{enumerate}

\smallskip

\noindent Below, we follow the ST13 solution procedure to obtain a slowly evolving mean-flow comprising solely of semigeostrophic frontogenesis. 

For convenience, we define an along-front velocity field $U_g$, which is in geostrophic balance with the horizontal pressure gradient 
\begin{align}
\label{geo_balance}
    {U}_g = -\frac{1}{f} \frac{\partial {P}}{\partial y}.
\end{align}
The buoyancy field $B$ is thus related to $U_g$ via the \textit{thermal-wind} balance,
\begin{align}
\label{eq_14}
\frac{\partial {U}_g}{\partial z} = \frac{S^2}{f},
\end{align}
where $S^2\equiv -\partial B/\partial y$.
Equations (\ref{eq_gov_a}) and (\ref{eq_gov_b}) can then be combined into a single equation, making use of (\ref{geo_balance}), 
\begin{align}
\label{gov_one}
    \frac{D^2U}{Dt^2} + (f^2 - \alpha^2) U = f^2 U_g.
\end{align}
To eliminate spontaneously emitted IWs, we apply a multiple scale approach and decompose all of the mean-flow fields into frontogenetic components (denoted by subscript $s$), which evolve over slow time scale $t_{s}\equiv\epsilon t$ {where $\epsilon=\alpha/f \ll1$}, and the spontaneously emitted IW components (denoted by subscript IW), which evolve over the fast time scale $t_{f}\equiv  t$, viz. 
\begin{subequations}
\label{decomp_all}
\begin{align}
\label{eq_ar11}
    U&= U_s(y,z,t_s) + \epsilon^{3/2}  U_\text{IW}(y,z,t_f) ,\\
\label{decomp_1}
   {V} &=\epsilon {V}_s(y,z,t_s) + \epsilon^{3/2} V_{\text{IW}}(y,z,t_f), \\
\label{decomp_1_1}
   {W} &=\epsilon{W}_s (y,z,t_s) + \epsilon^{3/2} W_{\text{IW}}(y,z,t_f), \\
\label{decomp_2}
    B &= B_s(y,z,t_s) + \epsilon^{3/2} B_{\text{IW}}(y,z,t_f) , \\
\label{decomp_3}
    P &= P_s(y,z,t_s) + \epsilon^{3/2} P_{\text{IW}}(y,z,t_f).
%
\end{align}
\end{subequations}
The $\mathcal{O}(\epsilon^0)$ terms in (\ref{decomp_all}) are associated with the geostrophic fields, and the $\mathcal{O}(\epsilon^1)$ terms are associated with the ASC. The small parameter $\epsilon$ illustrates that in semigeostrophic frontogenesis (e.g., HB72 model), the cross-front ASC is always weaker than the along-front geostrophic velocity. The $\mathcal{O}(\epsilon^{3/2})$ terms are associated with the spontaneously emitted waves, which are weaker than the ASC when the baroclinicity is sufficiently strong  ($f^2/S^2\ll 1$). The $1/2$ power is a result of the distinguished limit $f^3\sim \alpha S^2$ (Appendix \ref{filter_igw2}), which is satisfied in all of the solutions presented here and allows for a clean ordering separation.  
The fast and slow time scales are related via $t_s = \epsilon t_f$ such that 
\begin{align}
\label{eq_ar21}
    \frac{\partial}{\partial t} = \frac{\partial}{\partial t_f} + \epsilon \frac{\partial}{\partial t_s}.
\end{align}
Filtering out the spontaneously emitted waves amounts to truncating the asymptotic series in (\ref{decomp_all}) at $\mathcal{O}(\epsilon)$, plugging the truncated decomposition into equations 
(\ref{gov_eqs}$a-e$), and collecting terms of similar $\epsilon$ orders . It can be shown (Appendix \ref{filter_igw2}) that (\ref{gov_eqs}$a-e$) keep the same form albeit with total flow fields ($U,V,W,B,P$) replaced by the slow flow fields ($U_s, V_s, W_s, B_s, P_s$) and the time derivative $\partial/\partial t$ replaced by the slow time derivative $\epsilon \partial/\partial {t_s}$, such that  $D/Dt_s = \epsilon \partial/{\partial t_s} + (V_s-\alpha y) \partial/{\partial y} + W_s \partial/{\partial z}$. 
These modified evolution equations for the slowly evolving mean-flow, which are only valid for time scales of $\mathcal{O}(\alpha^{-1}) $, are solved semi-analytically in \S \ref{semi-anal} and numerically in \S \ref{numerical_model}. 
To lighten the notation in the remaining paper, we omit the subscript `$s$' from the mean-flow fields, and it is understood that we refer solely to the slowly varying (frontogenetic) part. 

\subsection{Semi-analytical solution and mean-flow initial conditions}
\label{semi-anal}
Next, we outline the procedure to obtain semi-analytical solutions for 2D semigeostrophic frontogenesis and pick initial conditions for the buoyancy ($B$) and along front velocity field ($U$) that ensure that IWs are not generated due to geostrophic adjustment. Finally, these semi-analytical solutions are used to validate the numerical solutions in \S\ref{results}.

{Equations (\ref{gov_eqs}$a-e$) materially conserve the PV, $q=(f\hat{z} + \nabla\times \bU)\cdot \nabla B$\footnote{note that because we assume hydrostatic balance the lateral derivatives of vertical velocity are neglected from the PV definition.}, where
the material derivative is defined in (\ref{mat_der}).} 
{Using \textit{generalized momentum} coordinates 
\begin{subequations}
\label{gen_coor}
\begin{gather}
 \label{gcm_y_1}
Y = \ee^{\alpha t} \Big( y - \frac{{U}}{f} \Big), 
\,\,\,\,\,\
Z = z, 
\,\,\,\,\,\
T = t,     
\tag{\theequation $a-c$} 
\end{gather}
\end{subequations}
the evolution PV can be written as
\begin{subequations}
\begin{gather}
\label{PV_cons}
\frac{D q}{D T} = 0, 
\,\,\,\,\,\
q = f \frac{\partial {B}}{\partial Z}
\Bigg[ 1 + \frac{1}{f} \ee^{\alpha t} \frac{\partial {U}}{\partial Y}
\Bigg]^{-1},
\tag{\theequation $a$,$b$}    
\end{gather}
\end{subequations}
with the generalized material derivative defined as
\begin{align}
\label{mat_div_gen}
\frac{D}{DT} \equiv \frac{\partial}{\partial T} + {W} \frac{\partial}{\partial Z}.
\end{align}
}

\noindent Semi-analytical solutions can be obtained for an initially uniform PV, which we pick to be $q_0 = f N^2$ ($N^2$ denotes a constant base-state stratification). With this choice, (\ref{PV_cons}) becomes
\begin{align}
\label{311}
    f \frac{\partial B}{\partial Z} - q_0 \Big(1 + \frac{1}{f} \frac{\partial U}{\partial Y} \Big) = 0.
\end{align}
Because the PV is uniform the mean-flow buoyancy field can be generally defined as 
\begin{align}
\label{buoy_field_gmc}
    {B}(Y, Z, T) = N^2 Z + \underbrace{{B}_g(Y) + \Delta B(Y, Z, T)}_{B'(Y,Z,T)},
\end{align}
where $B_g(Y)$ represents an imposed initial buoyancy distribution. We pick
\begin{align}
\label{eq_b0}
    B_g(Y) = \mathscr{B} \erf({Y/\lambda}),
\end{align}
{which describes a localized frontal zone with magnitude $|\mathscr{B}|$ and horizontal width $\lambda$}.
The variable $\Delta B$ represents the induced buoyancy variations due to frontogenesis (or any other process) that are required to ensure that the PV is conserved. As shown in Appendix \ref{analytical_sol}, a few algebraic manipulations from (\ref{gov_one}) lead to the following evolution equation that describes the evolution of both frontogenetic flow and
spontaneously emitted IWs,   
\begin{align}
\label{eq_123}
\frac{\partial^2 {U}}{\partial T^2} + (f^2 - \alpha^2) {U} + N^2 \ee^{2\alpha T} \int \int_0^Z \frac{\partial^2 {U}}{\partial Y^2} dZ^\prime dZ= - f \ee^{\alpha T} \frac{dB_g}{dY} \int dZ.
\end{align}
The above equation can be solved for the along front velocity $U$ using 
{Fourier transform in $Y$ and cosine transform in $Z$ (to satisfy the free-slip boundary conditions),} given a prescribed initial buoyancy field $B_g(Y)$ (e.g., (\ref{eq_b0})) and assuming that the full-depth integrated right-hand-side vanishes. This last assumption implies that the solution contains only the baroclinic component of the flow, which is sensible for
frontogenesis.
{The solution of the horizontal component of the ASC $(V)$ involves cosine modes in the vertical to satisfy the free-slip boundary conditions, while the vertical component of the ASC $(W)$ and ASC streamfunction (\textPsi) involve sine modes to satisfy the rigid-lid boundary conditions. The buoyancy anomaly $(\Delta B)$ involves sine modes to satisfy the Dirichlet conditions at the top and bottom boundaries. } 
Once $U$ (by using (\ref{mst13}) and (\ref{eq_124}) ) is known the remaining fields ($V,W,\Delta B$) can be determined from (\ref{eq_331}) and (\ref{eq_333}), and transformed back from the \textit{generalized momentum} coordinates using the Jacobian of transformation 
\begin{align}
\label{def_J}
    \mathcal{J} = \ee^{\alpha T} \Bigg( {1+\frac{1}{f} \ee^{\alpha T} \frac{\partial {U}}{\partial Y}} \Bigg)^{-1}.
\end{align}

\begin{figure}
\centering
\includegraphics[width=\linewidth]{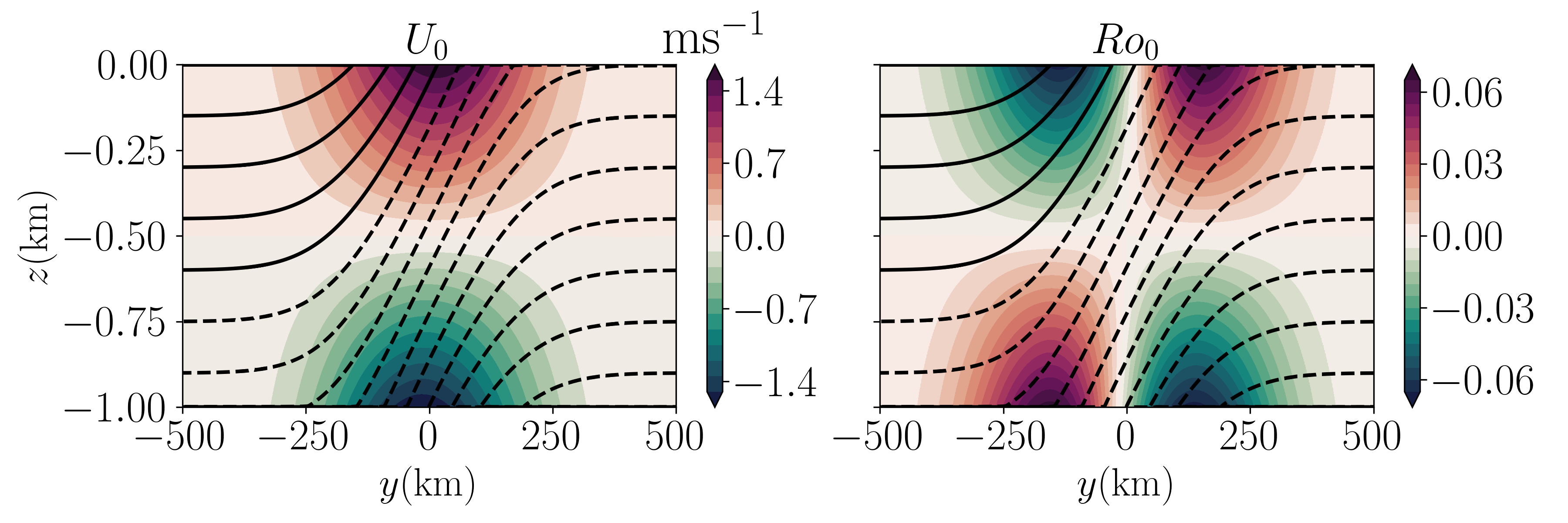}
\caption{The initial condition for $(a)$ the along front velocity $U(y,z,0)=U_{0}$ and $(b)$ Rossby number $(\Ro_0 =-\partial_y U_{0}/f)$. Contour lines of the initial buoyancy field $B(y,z,0)=B_0$ are displayed with a $0.012$ ms$^{-2}$ contour interval (solid and dotted line shows positive and negative values, respectively). The $N^2, \mathscr{B},$ and $\lambda$ values used in (\ref{buoy_field_gmc}) and (\ref{eq_b0}) are $10^{-2}$ s$^{-1}$, $-0.06$ m s$^{-2}$, and $200$ km, respectively. The methodology used to obtain $U_{0}$ and the initial buoyancy variation $\Delta B_0$ are discussed in Appendix \ref{analytic_ic}. The initial ASC is set to zero $(V_0=W_0=0)$ such that the initial conditions correspond to a geostrophically balanced front with a small root-mean-squared (rms) Rossby number (Ro$_\text{rms}\approx 0.03$).
}
\label{fig:figure1}
\end{figure}

In Appendix \ref{analytic_ic} we demonstrate how we use solutions to (\ref{eq_123}) with $\alpha=0$ to design initial conditions that ensure that there are no IW excitation due to geostrophic adjustment. 
{The initial buoyancy $B_0$ (e.g., (\ref{buoy_ss})) and along front velocity field $U_0$ (obtained from (\ref{ss_1})) used in our numerical model is shown in figure \ref{fig:figure1} ($V_0=W_0=0$)}.

Next, we filter out spontaneously emitted waves from the analytical solutions to match the numerical solutions. The same multiple-scale method that is used above (\ref{decomp_all}$a-e$) is applied to (\ref{eq_123}) yielding a waveless frontogenetic solution. The details of the resulting analytical model, which we call `modified' ST13, are discussed in Appendix \ref{filter_igw1} and compared with the original ST13 model and with HB72 model in figure \ref{fig:sol_cmp}.

\subsection{Internal wave evolution equations and initial conditions}
\label{sec_niw_mode}
To the slowly evolving mean-flow (\S \ref{mean_flow_config}), we add 2D hydrostatic, linear internal waves that evolve over the fast time scale $t_f$. 
The associated evolution equations for the IW velocity $(\bu=(u,v,w))$, buoyancy $(b)$, and pressure $(p)$ fields are
\begin{subequations}
\label{eq_wave_all}
\begin{align}
\label{eq_wave1}
\frac{Du}{Dt_f} + v \frac{\partial {U}}{\partial y} + w \frac{\partial {U}}{\partial z} - f v + \alpha u &= 0  \\
\label{eq_wave2}
\frac{Dv}{Dt_f} + v \frac{\partial {V}}{\partial y} + w \frac{\partial {V}}{\partial z} + f u - \alpha v &= -\frac{\partial p}{\partial y}  \\
\label{eq_wave3}
0 &= -\frac{\partial p}{\partial z} + b , \\
\label{eq_wave4}
\frac{Db}{Dt_f} + v \frac{\partial {B}}{\partial y} + w \frac{\partial {B}}{\partial z}  &= 0, \\
\label{eq_wave5}
\frac{\partial v}{\partial y} + \frac{\partial w}{\partial z} &= 0,
\end{align}
\end{subequations}
where $D/Dt_f=D/Dt$ (e.g. (\ref{mat_der})) with $\partial/\partial t$ replaced by $\partial/\partial t_f$, non-linear wave-wave terms are discarded.
To initialize vertical IW modes, we set $\alpha$ to zero and exploit the fact that the initial conditions for the mean-flow contain no ASC ($V_0=W_0=0$), such that  $D/Dt_f$ can be approximated by  $\partial/\partial {t_f}$. The problem configuration allows us to introduce a streamfunction $\psi$ such that
\begin{subequations}
\begin{align}
\label{def_v_w}
    v = \frac{\partial \psi}{\partial z}, 
    \,\,\,\,\,\,\,\,\,\,\,\
    w = -\frac{\partial \psi}{\partial y}, 
\tag{\theequation $a$,$b$} 
\end{align}
\end{subequations}
and simplify (\ref{eq_wave_all}$a-e$) to a single partial differential equation for $\psi$ (e.g., \citealp{whitt2013near})
\begin{align}
\label{eq_22}
    \Big (F^2 + \frac{\partial^2}{\partial t_f^2} \Big) \frac{\partial^2 \psi}{\partial z^2} + 2 S^2 \frac{\partial ^2 \psi}{\partial y \partial z} + N^2 \frac{\partial^2 \psi}{\partial y^2} = 0,
\end{align}
where $F^2 = f\Big( f - {\partial U}/{\partial y} \Big)$ and $S^2 = f \partial U/\partial z = - \partial B/\partial y$.
{To make progress we assume that the variables $F^2$ and $S^2$ are constant, which is justified in the frontal zonal (e.g., figure \ref{fig:figure1}, $-200 \text{km}<y<200\text{km}$) when the initial Rossby number is sufficiently small.}
Following \cite{gerkema2005near} we look for plane-wave solutions of the form
\begin{align}
\label{ansatz}
{\psi}(y,z) = \widetilde{\psi}(z) \exp{\{\ii l (y-A_1 z)\}}\exp{(-\ii \omega t_f}),
\end{align}
where $\omega$ is the wave frequency and $l$ is the wavenumber in the $y$-direction. Substituting the ansatz (\ref{ansatz}) into (\ref{eq_22}) yields
\begin{align}
\label{eq_eigen}
    \frac{d^2 \widetilde{\psi}}{dz^2} + l^2(A_1^2 - A_0) \widetilde{\psi} = 0,
\end{align}
where $A_0=N^2/(ff_\text{eff}-\omega^2)$, $A_1=S^2/(ff_\text{eff}-\omega^2)$ are constant, and $f_\text{eff} = f- \partial U/\partial y$ is the effective Coriolis frequency. Equation (\ref{eq_eigen}), subject to the boundary conditions,
\begin{align}
\label{eq_bcs}
    \widetilde{\psi}(z=0) = \widetilde{\psi}(z=-H)=0.   
\end{align}
has solutions of the form
\begin{align}
\widetilde{\psi}_n(z) = C_n \sin (m_n z), \,\,\,\,\,\ m_n = \frac{n\upi}{H},
\end{align}
where $m_n$ is the vertical wavenumber of the $n$th mode, and $C_n\in\mathbb{R}$.
The remaining fields are given by ((\ref{def_v_w}), (\ref{eq_wave1}) and (\ref{eq_wave4}))
\begin{subequations}
\label{IW_fields}
\begin{align}
\label{IW_fields_a}
[v,w]&=[\underbrace{-\ii l A_1 \widetilde{\psi}_n+\partial{\widetilde{\psi}_n}/{\partial z}}_{\widetilde{v}},\underbrace{-\ii l  \widetilde{\psi}_n}_{\widetilde{w}}]\exp{\{\ii l (y-A_1 z)-\ii\omega t_f\}},
\end{align}
\begin{align}
\label{u_eignfun}
   {u} &= \frac{\ii}{\omega f} \big( f f_\text{eff} {\widetilde{v}} -S^2 {\widetilde{w}} \big)\exp{\{\ii l (y-A_1 z)-\ii\omega t_f\}},
\end{align}
\begin{align}
\label{b_eigenfun}
 {b} &= \frac{\ii}{\omega} \big( S^2 \widetilde{v} - N^2 \widetilde{w} \big)\exp{\{\ii l (y-A_1 z)-\ii\omega t_f\}},
 \end{align}
\end{subequations}
with the dispersion relation
\begin{align}
\label{c11}
  \omega^2 = f f_\text{eff}+ \frac{l^2 N^2}{2 m_n^2} \Bigg[ 1 \pm \sqrt{1 + \frac{4 m_n^2 S^4}{l^2 N^4}} \Bigg] \nonumber \\
=  f f_\text{eff} + \frac{l^2 N^2}{2 m_n^2} \Bigg[ 1 \pm \sqrt{1 + \frac{4}{\Ri_g \Bu}} \Bigg],
\end{align}
where $\Ri_g=N^2/(\partial_z U)^2 = f^2 N^2/S^4$ is the Richardson number of the along-front geostrophic flow, and $\Bu=(l^2 N^2)/(f^2 m_n^2)$ is the IW Burger number. For finite scale low-mode IWs in a geostrophically balanced frontal zone $\Bu \sim \mathcal{O}(1), \Ri_g\gg 1$ and $4/(\Ri_g \Bu) \ll 1$. The expression in the squared root can thus be expanded in a Taylor series, and (\ref{c11}) becomes 
\begin{align}
\label{approx_disp}
    \omega^2 \approx f f_\text{eff} + \frac{l^2 N^2}{2 m_n^2} \Bigg[ 1 \pm \Bigg( 1 + 2 \frac{m_n^2 S^4}{l^2 N^4} \Bigg) \Bigg].  
\end{align}
From (\ref{approx_disp}) the minimum IW frequency $\omega_\text{min}$ is  
\begin{align}
    \label{min_freq}
    \omega_\text{min} \approx 
    f \sqrt{1 + Ro - \frac{1}{Ri_g}},  
\end{align}
which recovers the expression derived in \cite{whitt2013near}, and shows that vorticity and baroclinicity allow the IW frequency to be lower than the inertial frequency.\footnote{The imposed geostrophic strain modifies the IW frequency at $\mathcal{O}((\alpha/f)^2)$, which is negligible in our case compared with the effects of vorticity and baroclinicity \citep{jing2017energy}. } The corresponding horizontal group velocity $c_{g_y}$ takes the form 
\begin{align}
\label{cg_min_freq}
    c_{g_y}  \approx \pm \frac{N^3 (\omega^2 - \omega^2_\text{min})^{3/2}}{m_n \omega(2S^4+ N^2(\omega^2-f f_\text{eff}))},
\end{align}
where the positive (negative) sign corresponds to $l>0$ ($l<0$).

\begin{figure}
\centering
\includegraphics[width=\textwidth]{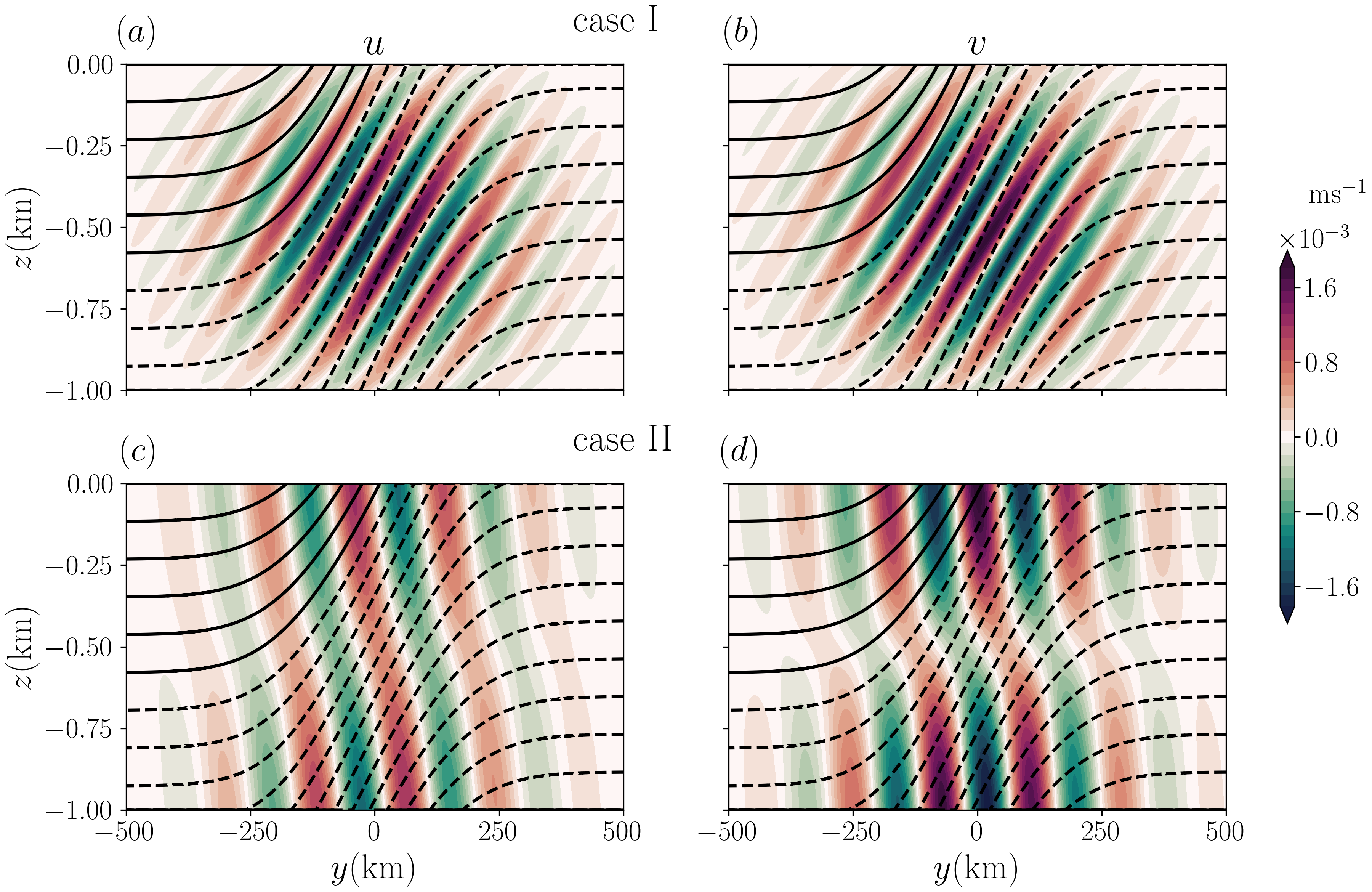}
\caption{Initial conditions for mode-$1$ $(a,c)$ along-front and $(b,d)$ cross-front IW velocities. The buoyancy contour lines are displayed with a $0.012$ ms$^{-2}$ contour interval (solid and dotted line shows positive and negative values, respectively). Panels ($a$,$b$) correspond to IW phase lines that are parallel to buoyancy contour lines near the frontal zone (Case I - NIWs), whereas panels ($c$,$d$) correspond to IW phase lines that are tilted against the buoyancy contour lines near the frontal zone (Case II - high-frequency IWs, $\omega =1.5f$).
}
\label{fig:wave_sol}
\end{figure}

Our IW initial conditions consist of a Gaussian packet of a mode-$1$ IW ($m_1$ in (\ref{IW_fields}$a-c$)), with {a horizontal width of three wavelengths ($6\pi/l$}), and with phase lines approximately parallel to (case I) or tilted against (case II) isopycnals (figure \ref{fig:wave_sol}). As discussed in T12, case I corresponds to a minimum frequency IW (NIW) that is phase-locked to the frontal zone ($ c_{g_y} \to 0$, (\ref{cg_min_freq})) and is therefore more likely to exchange energy with the frontal circulations. 
On the contrary, case II corresponds to a higher frequency IW (namely, $\omega =1.5f$) that can propagate away from the frontal zone ($ c_{g_y} \ne 0$, (\ref{cg_min_freq})) and is therefore less likely to exchange energy with the frontal flow.
 In \S \ref{sec_energetics} we compare and contrast between the two cases.

\section{Internal wave energy equations}
\label{sec_wave_ke}
The inviscid IW KE equation is obtained by multiplying (\ref{eq_wave1}) with $u$, (\ref{eq_wave2}) with $v$, and taking account of the numerical dissipation
(e.g., (\ref{diffusive_op})) leading to 
\begin{align}
\label{wave_ke_inviscid}
\frac{D \mathcal{K}}{D t_f}  & =
\underbrace{-u w \frac{\partial U}{\partial z}}_{\text{GSP}}
\underbrace{- u v \frac{\partial U}{\partial y}}_{\text{LSP}}
\underbrace{- \alpha(u^2 - v^2)}_{\text{DSP}}
\underbrace{- \delta v^2}_{\text{CP}}  \nonumber 
\underbrace{- v w \frac{\partial V}{\partial z}}_{\text{AGSP}}  
\underbrace{+ w b}_{\text{BFLUX}} \\
&\underbrace{- \nabla \cdot \mathbf{v} p}_{\text{PWORK}},
\end{align}
where $D/Dt_f=\partial/\partial t_f + (V-\alpha y)\partial/\partial y+W\partial/\partial z$, 
$\mathcal{K}= 1/2(u^2+v^2)$, $\nabla = \hat{y} \partial/\partial y  + \hat{z} \partial/\partial z $, $\mathbf{v} = \hat{y} v + \hat{z} w$ and $\delta=\partial V/\partial y$ is the horizontal divergence associated with the ASC. 


The terms in (\ref{wave_ke_inviscid}) are the geostrophic-shear-production (GSP), denoting wave-mean flow energy exchanges associated with the geostrophic vertical shear; the lateral-shear-production (LSP),  denoting wave-mean flow  energy exchanges associated with the geostrophic lateral shear; the deformation-shear-production (DSP), denoting energy exchanges due to the imposed deformation flow; the ageostrophic-shear-production (AGSP), denoting wave-mean flow energy exchanges associated with the ageostrophic vertical shear; the convergence-production (CP), denoting wave-mean flow energy exchanges associated with the lateral ageostrophic divergent motions; the buoyancy flux (BFLUX), indicating the energy exchanges between wave kinetic and potential energies (see (\ref{wave_pe_inviscid})); and the pressure work (PWORK), denoting wave energy changes due to the propagation of pressure perturbations. 

The inviscid IW potential energy (PE) equation is obtained by multiplying (\ref{eq_wave4}) with $b/N^2$ (assuming that $\partial B/\partial z \approx N^2$ is a constant), leading to 
\begin{align}
\label{wave_pe_inviscid}
    \frac{D \mathcal{P}}{D t_f} = 
    - \frac{vb}{N^2} \frac{\partial B}{\partial y} -\underbrace{ w b}_{\text{BFLUX}},
\end{align}
where $\mathcal{P}=1/2 (b^2/N^2)$. The first term on the right-hand-side of (\ref{wave_pe_inviscid}) is generally smaller than the KE equation terms and is therefore not shown in the analysis that follows. 


In \S \ref{sec_energetics} we evaluate all the terms in (\ref{wave_ke_inviscid}) in several numerical experiments with different strain magnitudes, IW initial conditions, and vertical modes. 
In some cases the terms in (\ref{wave_ke_inviscid}) are domain averaged and time integrated,  where the notation 
$\langle \cdot \rangle$ denotes the domain average.
Note that $\langle \text{PWORK} \rangle = 0$ for our choice of boundary conditions (\S \ref{numerical_model}). We further introduce the notation $\Delta \langle  \mathcal{K} \rangle (t;t_0) = \langle \mathcal{K} \rangle (t)- \langle \mathcal{K} \rangle(t_0)$, to denote the change in domain averaged wave KE at time $t$ relative to another time $t_0$. 
\section{{Numerical setup}}
\label{numerical_model}
\begin{table}
  \begin{center}
\def~{\hphantom{0}}
  \begin{tabular}{lccc}
      Parameter  & Definition  &  Value \\[3pt]
       $L$ & domain size in $y$-direction & $1000$ km ($600$ km after nesting) \\
       $H$ & domain size in $z$-direction & $1$ km \\
       $\mathsf{N}_y$ & number of uniform grid points in $y$-direction & $2000$ \\
       $\mathsf{N}_z$ & number of Chebyshev points in $z$-direction & $240$ \\
       $\mathscr{B}$ & amplitude of the front & $-0.06$ m s$^{-2}$ \\
       $\lambda$ & cross-front length scale & $200$ km \\
       $N$   & buoyancy frequency & $10^{-2}$ s$^{-1}$ \\
       $f$   & Coriolis frequency & $10^{-4}$ s$^{-1}$ \\
       $Ro_\text{rms}$ & Rossby number of front & 0.03 \\
       $Ri_g$ & Richardson number of the front & 15 \\
       $\alpha$ & geostrophic strain & $(0.04f, 0.1f, 0.25f)$ s$^{-1}$ \\
       $\omega$ & frequency of NIW & $0.97 \times 10^{-4}$ s$^{-1}$ \\
       $l$ & horizontal wavenumber of mode-$1$ NIW & $-4.7 \times 10^{-5}$ m$^{-1}$ \\ 
       $Bu$ & Burger number of mode-$1$ NIW & 2.24 \\
       $\nu$ & viscosity & $2 \times 10^{-4}$ m$^2$ s$^{-1}$\\
       $\kappa$ & diffusivity & $2 \times 10^{-4}$ m$^2$ s$^{-1}$\\
       $\nu_h$ & hyperviscosity & $10^8$ m$^4$ s$^{-1}$\\
       $\kappa_h$ & hyperdiffusivity & $10^8$ m$^4$ s$^{-1}$\\
       $\sigma_y$  & $y$ tapering scale & $3$ km\\
        $\sigma_z$  & $z$ tapering scale & $2$ m ($\sim 5$ grid points)\\
  \end{tabular}
  \caption{Description of the simulation parameters. 
  }
  \label{tab:params}
  \end{center}
\end{table}
The problem configuration detailed in \S \ref{gov_eqs} describes a quasi-linear model of a slowly-evolving mean-flow and a fast-evolving IW vertical mode. In the numerical model, the mean flow buoyancy field $B$ is defined as $B(y,z,t)=N^2 z+ B^\prime(y,z,t)$, where $B^\prime$ comprises $B_g$ and $\Delta B$ (e.g. \ref{buoy_field_gmc}). Because $N^2$ is prescribed we only solve for $B^\prime$, which represents buoyancy variation due to  frontogenesis.
We solve the following mean-flow equations numerically (see  \ref{num_eqs}$a-e$)
\begin{subequations}
\label{num_eqs_1}
\begin{align}
    \frac{D U}{Dt_\text{s}} - fV + \alpha U &= \mathscr{D}(U),  \\  
    \frac{D V}{Dt_\text{s}} - \alpha V &= \mathcal{T}_1 fU  -\frac{\partial P}{\partial y} + \mathscr{D}(V), \\
    0 &= -\frac{\partial P}{\partial z} + B^\prime, \\
    \frac{D B^\prime_\text{s}}{D t} + N^2 W &= \mathscr{D}(B^\prime), \\
    \frac{\partial V}{\partial y} + \frac{\partial W}{\partial z} &= 0,
\end{align}
\end{subequations}
with the material derivative  
\begin{align*}
 \frac{D}{Dt_\text{s}}  = \epsilon \frac{\partial}{\partial t} + (V-\alpha y) \frac{\partial}{\partial y} + W \frac{\partial}{\partial z},
\end{align*}
where $\epsilon=\alpha/f$.
The diffusivity operator 
\begin{align}
\label{diffusive_op}
    \mathscr{D} \equiv \nu \Big( \frac{\partial^2}{\partial y^2} + \frac{\partial^2}{\partial z^2} \Big) - \nu_h \frac{\partial^4  }{\partial y^4},
\end{align}
is added to ensure numerical stability, and $\mathcal{T}_1$ is the tapering function
\begin{align}
\label{taper_y}
  \mathcal{T}_1(y) = 1 - \ee^{-((y+L/2)/\sigma_y)^2} - \ee^{-((y-L/2)/\sigma_y)^2},
\end{align}
which is added to the Coriolis term in the $y$-momentum equation to ensure no pressure gradients develop near the vertical walls \citep{winters2012modelling}. 

The following wave equations are solved numerically (see \ref{eq_wave_all}$a-e$)
\begin{subequations}
\label{eq_wave_all_1a}
\begin{align}
\label{eq_wave1a}
\frac{Du}{Dt_f} + v \frac{\partial {U}}{\partial y} + w \frac{\partial {U}}{\partial z} - f v + \alpha u &= \mathscr{D}(u) + \mathcal{F}_u, \\
\label{eq_wave2a}
\frac{Dv}{Dt_f} + v \frac{\partial {V}}{\partial y} + w \frac{\partial {V}}{\partial z} - \alpha v &= \mathcal{T}_1 f u -\frac{\partial p}{\partial y} + \mathscr{D}(v) + \mathcal{F}_v,\\
\label{eq_wave3a}
0 &= -\frac{\partial p}{\partial z} + b, \\
\label{eq_wave4a}
\frac{Db}{Dt_f} + v \frac{\partial {B}}{\partial y} + w \frac{\partial {B}}{\partial z}  &= \mathscr{D}(b) + \mathcal{F}_b, \\
\label{eq_wave5a}
\frac{\partial v}{\partial y} + \frac{\partial w}{\partial z} &= 0,
\end{align}
\end{subequations}
with the material derivative  
\begin{align*}
 \frac{D}{Dt_f} = \frac{\partial}{\partial t} + (V-\alpha y) \frac{\partial}{\partial y} + W \frac{\partial}{\partial z}.  
\end{align*}
The diffusivity operator $\mathscr{D}$ is given in (\ref{diffusive_op}) and leads to the following numerical wave KE dissipation
\begin{align}
\label{num_dissip}
\text{DISP}=u\mathscr{D}(u)+ v\mathscr{D}(v).
\end{align}
In the analysis that follows DISP is time integrated and domain averaged, as discussed  in  \S \ref{sec_wave_ke}.

In the current manuscript, we assume that the mean-flow and IW fields are decoupled. 
In a forthcoming publication we allow the IWs to feedback on the mean flow using a phase-averaging operator over the fast-evolving IWs, which leads to the inclusion of averaged IW fluxes in the mean-flow evolution equations (i.e., the full quasilinear model).

The terms $(\mathcal{F}_u, \mathcal{F}_v, \mathcal{F}_b) \equiv 1/2(\alpha u, \alpha v, \alpha b)$ are added on the right-hand side of the wave momentum and buoyancy equations to ensure numerical energy conservation.
This is because the equations we solve for numerically are invariant in the $x$ direction and, consequently, the imposed strain field is divergent and results in a wave KE and PE sink at a rate that equals $\alpha \mathcal{K}$ and $\alpha \mathcal{P}$, respectively. The IW energy conservation and the maintenance of the IW amplitude is particularly important in the full quasilinear model. Previous studies (e.g., asymptotic theory of \cite{xie2015generalised}, and idealized numerical simulation of \cite{thomas2021forward,xie2020downscale}) showed that the relative magnitude of the IWs compared with that of the mean flow is a key parameter controlling the wave-mean flow interactions. Therefore, we do not want the advection by the imposed deformation flow to remove energy from the domain and rapidly decrease the wave amplitude. We make sure that the results presented in the current manuscript are unaffected by these terms (Appendix D).

{The boundary conditions for velocity (both mean flow and IW) are free-slip walls in the horizontal direction and free-slip rigid lid in the vertical direction. Dirichlet boundary conditions are used for both mean-flow and wave buoyancy perturbations in the top and bottom boundaries (specifically, $B^\prime =b=0$ at $z=0,-H$). These specified boundary conditions are identical to the ones used by the semi-analytical solutions of \cite{shakespeare2013generalized} (\S \ref{semi-anal} and Appendix B).  The initial conditions for the mean flow and the IWs are discussed 
in \S \ref{gov_eqs} (figures \ref{fig:figure1} and \ref{fig:wave_sol}).  The initial buoyancy fields $B^\prime$ and $b$ are tapered to zero using a similar tapering function to (\ref{taper_y}), i.e., $\mathcal{T}_2(z)=1 - e^{-(z/\sigma_z)^2} - e^{-((z+H)/\sigma_z)^2}$, to ensure that the Dirichlet boundary conditions are met. }

The mean-flow and IW evolution equations above are solved using the pseudo-spectral code Dedalus \citep{burns2020dedalus} for three different values of imposed geostrophic strain $\alpha=0.04f, 0.1f$ and $0.25f$. All fields are expanded with Chebyshev polynomials in the vertical direction and with cosine/sine expansions in the horizontal direction, with a $3/2$ de-aliasing factor. Time-stepping is performed using a third-order $4$-step implicit-explicit Runge-Kutta scheme with a time-step of $20$s. Details of the simulation parameters are given in table \ref{tab:params}.
 
During the later stage of frontogenesis (i.e., the \textit{superexponential} stage), the frontal width rapidly decreases, thus requiring smaller grid spacing for adequate resolution. To this end, we add a nest to the original (`parent') grid once the front enters the \textit{superexponential} stage and restart the numerical integration using the `parent' values for both the mean-flow and IW variables as initial conditions. These new initial conditions are interpolated to the nested grid, which contains the same number of grid points as the `parent' grid (table \ref{tab:params}) but with a smaller horizontal domain size ($\pm 300$km), leading to a decrease in the horizontal grid spacing from $500$m to $300$m. 


\section{Mean-flow evolution}
\label{results}

\begin{figure}
\centering
\includegraphics[width=\linewidth]{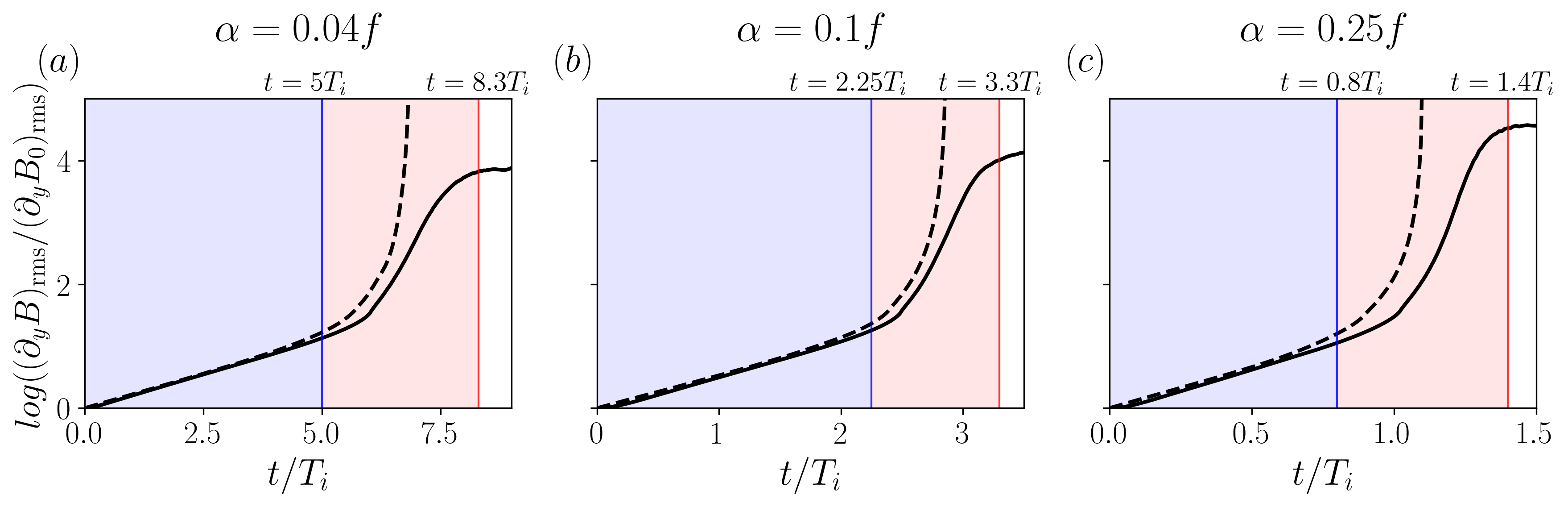}
\caption{The root-mean-squared (rms) horizontal buoyancy gradient evolution $(\partial_y B)_{\text{rms}}$ 
computed in the frontal region using the semi-analytical (dashed) and numerical solutions (solid), for three different values of $\alpha$.
The frontal region is identified as the region where $(\partial_y B)^2 > 0.1 (\partial_y B)_{\text{max}}^2$. The end of the \textit{exponential} and \textit{superexponential} frontogenetic stages are marked above the thin vertical blue and red lines, respectively. Time is normalized by the inertial period $T_i$. 
}
\label{fig:buoy_cmp}
\end{figure}

\begin{figure}
\centering
\includegraphics[width=\linewidth]{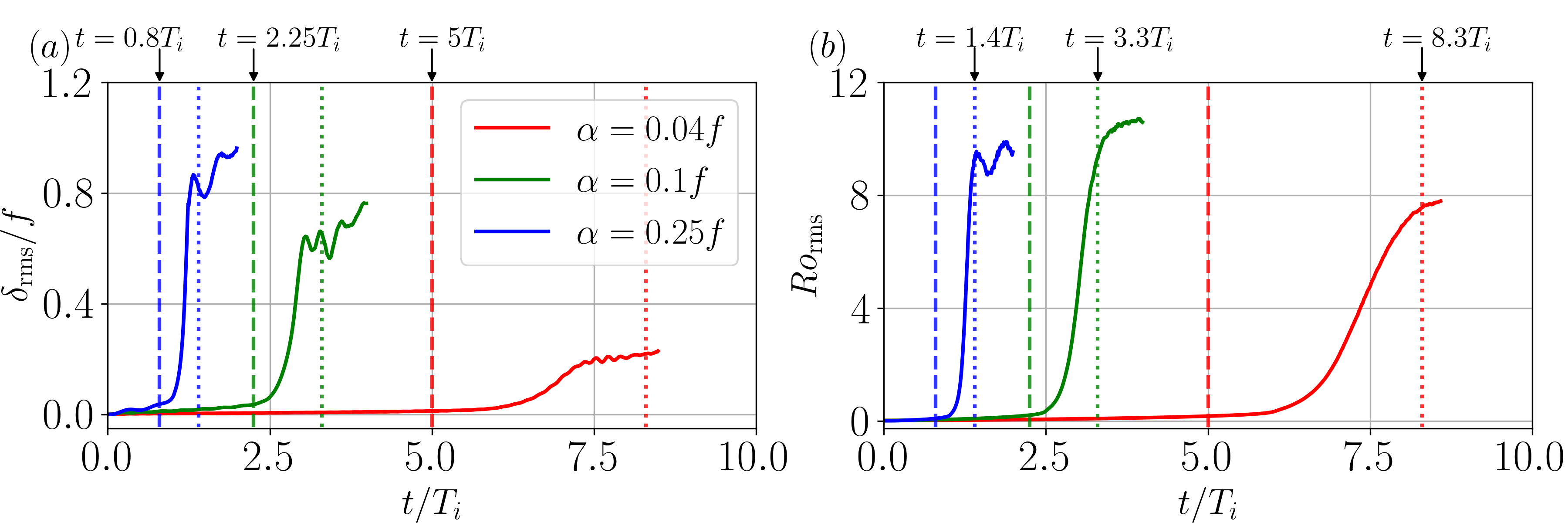}
\caption{The rms normalized divergence ($\delta_{\text{rms}}/f$) evolution $(a)$ and vorticity ($\Ro_{\text{rms}}\equiv \zeta_{\text{rms}}/f$) evolution $(b)$, computed in the frontal region {(same definition as in figure \ref{fig:buoy_cmp})} for three different values of $\alpha$. Dashed (dotted) vertical lines indicate the end of the \textit{exponential} (\textit{superexponential}) frontogenesis stages. Time is normalized by the inertial period $T_i$. 
}
\label{fig:div_ro_evol}
\end{figure}

In 2D semigeostrophic frontogenesis (HB72, \citealp{hoskins1982mathematical}), the initial frontal sharpening is dominated by the externally imposed geostrophic strain field $\alpha$, leading to an exponential sharpening rate (the \textit{exponential} stage). The convergent ageostrophic secondary circulation (ASC) that develops about the front gradually becomes stronger until it dominates the geostrophic strain, driving a superexponential sharpening rate that leads to a finite time singularity in the inviscid limit (the \textit{superexponential} stage). These two growth stages are shown in figures \ref{fig:buoy_cmp} and \ref{fig:div_ro_evol} for different values of  $\alpha$. {The rms of the horizontal buoyancy gradient evolution ($(\partial_y B)_\text{rms}$)  averaged over the frontal region} shows a good agreement between the numerical and analytical values, particularly during the exponential stage (figure \ref{fig:buoy_cmp}). The difference between the analytical and numerical solutions during the superexponential stage is because numerical diffusion (\ref{diffusive_op}) acts to halt the frontal sharpening before the finite-time singularity is reached. Moreover, a comparison across different $\alpha$ values (figure \ref{fig:buoy_cmp}$(a-c)$) shows that as $\alpha$ increases, the sharpening rate also increases, and the duration of the exponential stage is shortened. Accordingly, the analytical frontogenesis duration (until a finite-time singularity is reached) reduces from 7 to 1.1 inertial periods as $\alpha$ increases from $0.04f$ to $0.25f$. 

Strong buoyancy gradients at frontal regions are often associated with strong divergence $\delta=\partial_y V$ and vorticity $\zeta=-\partial_y U$ signals. Indeed the rms values of these quantities, averaged over the frontal region, increase rapidly as the front sharpens (figure \ref{fig:div_ro_evol}). Largely consistent with semigeostrophic frontogenesis, the simulated frontal flow is characterized by $\delta_{\text{rms}}/f \le \alpha/f$ and ${\Ro}_{\text{rms}}={\zeta}_{\text{rms}}/f \le 1$ during the exponential phase (dashed vertical lines in figure \ref{fig:div_ro_evol}), whereas during the superexponential stage it is characterized by $\delta_{\text{rms}}/f \gg \alpha/f $ and ${\Ro}_{\text{rms}}\gg 1$. In addition, $\text{Ro}_{\text{rms}}\gg  \delta_{\text{rms}}/f $ at all times (the along front geostrophic velocity is always larger than the cross front ageostrophic velocity) with final $\delta_{\text{rms}}/f $ values approaching $\mathcal{O}(1)$ for the cases with stronger strain values. 

\begin{figure}
    \centering
    \includegraphics[width=\textwidth]{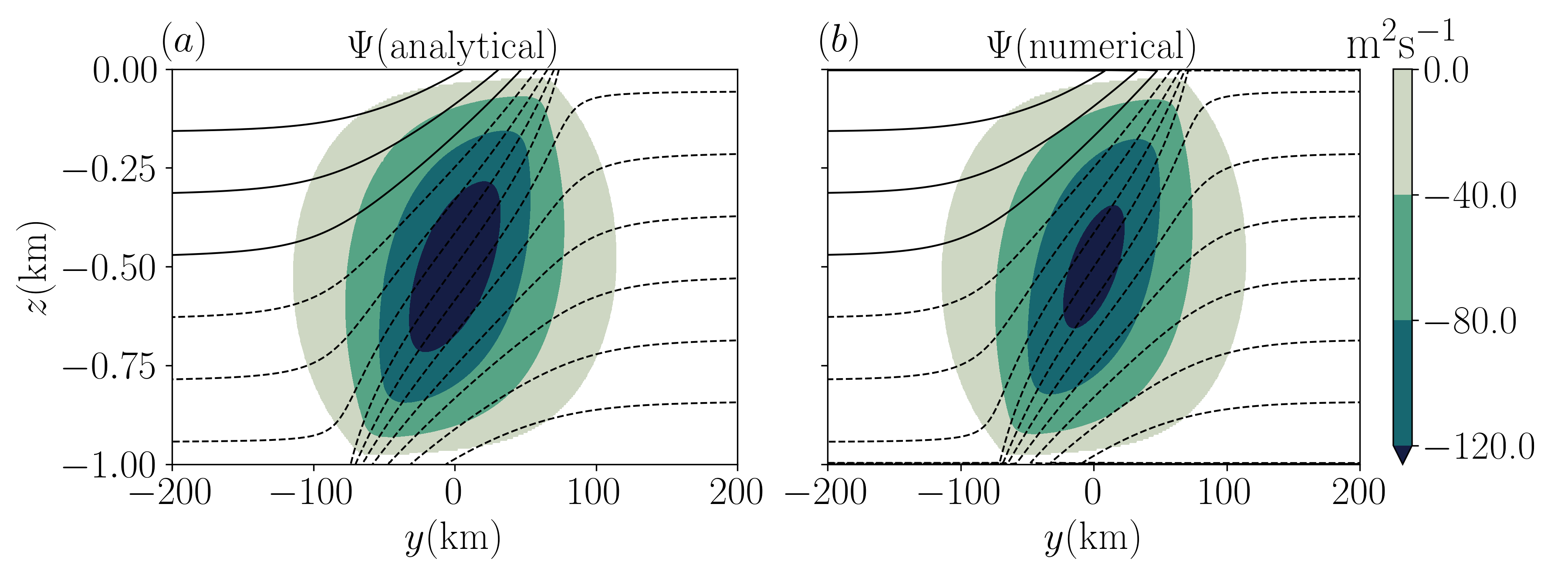}
    \caption{Snapshots of the ASC streamfunciton $\text{\textPsi}$ computed using the analytical $(a)$ and numerical $(b)$ solutions for $\alpha=0.1f$, during the \textit{exponential} frontognesis stage ($t=2T_i$, figure \ref{fig:buoy_cmp}($b$)). The associated buoyancy contour lines are displayed with a $0.016$ ms$^{-2}$ contour interval (solid and dotted line shows positive and negative values, respectively).  
    }
    \label{fig:strm_cmp}
\end{figure}

\begin{figure}
    \centering
    \includegraphics[width=\textwidth]{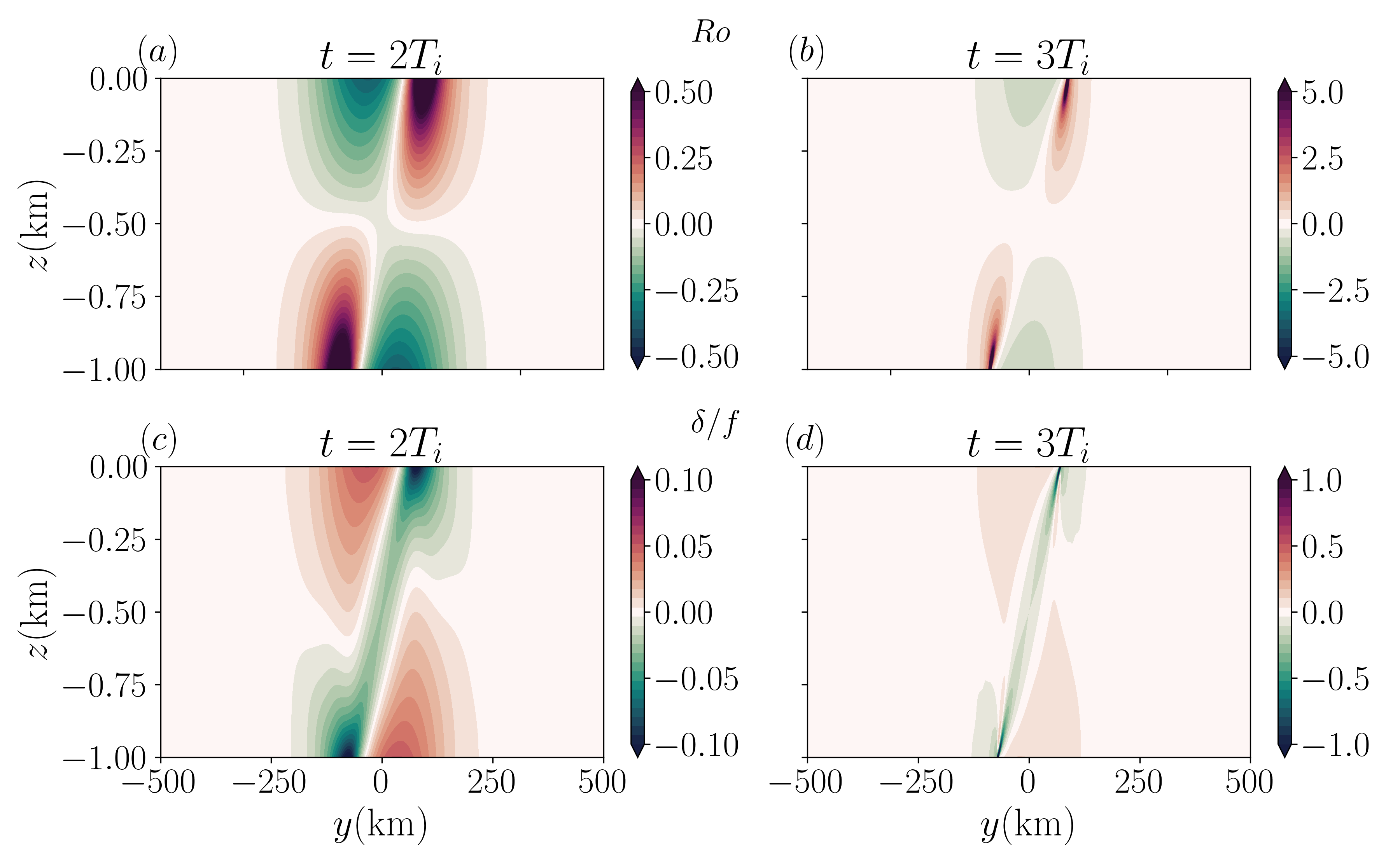}
    \caption{Snapshots of $(a,b)$ the Rossby number $\Ro$ and $(c,d)$ the normalized horizontal divergence of the ASC $\delta/f$, in the numerical simulation with $\alpha=0.1f$. Panels $(a,c)$ and $(b,d)$ correspond to the \textit{exponential} and \textit{superexponential} frontogenesis stages, respectively (see figure \ref{fig:buoy_cmp}$(b)$). $T_i$ is the inertial period. Note the different colorbar ranges between panels $a$ and $b$, and $c$ and $d$.
    }
    \label{fig:vor_cmp}
\end{figure}

A snapshot of the ASC streamfunction $\text{\textPsi}$ during the late \textit{exponential} stage (blue shaded region in figure \ref{fig:buoy_cmp}(b)) shows a good match between the semi-analytical and the numerical solutions (figure \ref{fig:strm_cmp}). As expected, the ASC is clockwise, leading to an energy conversion from APE to KE and restratification \citep{mcwilliams2016submesoscale,barkan2019role}.
The isopycnals are closer together near the top (bottom) boundaries at $y=70$ km ($y=-70$), where frontogenesis is strongest. At these locations the ASC is convergent ($\delta/f<0$; figure \ref{fig:vor_cmp}c) and the geostrophic vorticity is cyclonic ($\Ro>0$; figure \ref{fig:vor_cmp}a). Outside of these strong frontogenetic regions the flow is characterized by weaker divergence ($\delta/f>0$) and anticyclonic vorticity $(\Ro<0)$. 

One inertial period later, in the \textit{superexponential} stage, the asymmetry between cyclonic/convergent and anticyclonic/divergent circulation is enhanced, with near-surface convergence and vorticity values that increase by an order of magnitude. During that time, the frontal width (computed in the region where $(\partial_y B)^2 > 0.1(\partial_y B)_{\text{max}}^2$, at $z=-10$ m ) is decreased from around $120$ km to $5$ km. 

%
\section{Energy exchanges}
\label{sec_energetics}
The front-IW energy exchanges are explored for minimum frequency (near-inertial) and high-frequency waves (cases I and II in figure \ref{fig:wave_sol}) with vertical modes 1-3 and subject to three different imposed strain values.  We distinguish between energy exchanges during the \textit{exponential} and \textit{superexponential} frontogenetic stages (figure \ref{fig:buoy_cmp}), which are characteristic of oceanic mesoscale and submesoscale frontogenesis, respectively \citep{barkan2019role}.

The phase structure and KE exchanges with the mean flow are substantially different between near-inertial and high-frequency waves (figure \ref{fig:wave_vel} and supplementary movie 1). The NIW remains in the frontal zone because the horizontal group velocity $ c_{g_y} \to 0$ (\ref{cg_min_freq}), as discussed in T12, and nearly all of its kinetic-energy remains in the frontal zone during frontogenesis (figure \ref{fig:wave_vel}($a$)). This suggests that NIWs are likely to exchange energy with the front. On the contrary, the high-frequency wave is able to escape the frontal zone ($ c_{g_y} \ne 0$) with nearly all of its energy found outside the frontal region before the \textit{superexponential} stage is reached (figure \ref{fig:wave_vel}($d$)). The higher the initial IW frequency is, the faster it escapes the frontal region (the intersection between the solid and dashed magenta lines in figure \ref{fig:wave_vel}$(d)$ is shifted to the left by $\approx 0.2 T_i$ when the initial IW frequency is increased from $1.5f$ to $3f$).
\subsection{Case I: minimum frequency wave (NIW)}
\label{sec_min_freq}
The dominant KE exchange terms (\ref{wave_ke_inviscid}) for the minimum frequency wave include the DSP and AGSP (figure \ref{fig:wKE}). As discussed in T12 and \cite{whitt2013near}, the NIW is able to extract energy from the imposed deformation field ($\text{DSP} > 0$) when the frontal baroclinicity and vorticity modify the wave polarization relations, leading to rectilinear hodographs (i.e., $|v| > |u|$; figure\ref{fig:wave_vel}($c,d$)) and anisotropic horizontal momentum fluxes.\footnote{specifically for a minimum frequency wave $|u|/|v| \approx (1+Ro-Ri_g^{-1})^{1/2}$, as discussed in \cite{whitt2013near}.} The NIW loses its energy to the ASC when the wave isophases are tilted with the agostrophic shear ({solid blue line in figure \ref{fig:wave_vel}$(d)$}; $\text{AGSP}<0$). Similarly to T12, the AGSP is the main inviscid mechanism that drains  NIW KE. Quantitatively, both the DSP and AGSP remain largely unchanged when integrated separately over the \textit{exponential} and \textit{superexponential} stages, for all simulated $\alpha$ values (table \ref{tab:table1}).

The convergence production ($\text{CP}\equiv -\delta v^2$) is a newly identified mechanism for IW-front energy exchanges, which is associated with the convergence (or divergence) of the ASC. Convergent (divergent) regions correspond to $\text{CP}>0\, (\text{CP}<0)$ and wave KE gain (loss). This particular energy exchange mechanism is absent in T12, and the QG-NIW theories \citep{xie2015generalised,rocha2018stimulated,thomas2020near}, where the balanced (frontal) flow is horizontally non-divergent. 

During the \textit{exponential} stage, the convergence of the ASC in the frontal (cyclonic) region is rather weak, and is comparable to the divergence of the ASC in the anticyclonic region (figure  \ref{fig:vor_cmp}$(c)$). As a result, there is a cancellation when CP is domain-averaged, leading to small values compared with $\langle \text{DSP} \rangle$ (red and blue line in figure \ref{fig:wKE}, blue shading). During the \textit{superexponential} stage however, when frontal sharpening is primarily driven by the convergence of the ASC ($|\delta| \sim \mathcal{O}(f)$; figure \ref{fig:vor_cmp}$(d)$), CP gradually begins to dominate the NIW KE gain (red and blue lines in figure \ref{fig:wKE}, red shading). Quantitatively, when integrated over the \textit{superexponential} stage only, $\langle \text{CP} \rangle >\langle \text{DSP} \rangle$ for all simulated strain values (table \ref{tab:table1}). This CP dominance is particularly evident when the DSP and CP terms are averaged separately inside and outside the frontal zone (denoted by `F' and `OF', respectively; figure \ref{fig:fig_x}).

Inside the frontal zone, the time integrated $\langle \text{CP} \rangle_F$ increases rapidly during the \textit{superexponential} stage, coinciding with the rapid convergence increase of the ASC (figures \ref{fig:div_ro_evol}$(a)$ and \ref{fig:vor_cmp}$(d)$), and dominates $\langle \text{DSP} \rangle_F$  (solid red and blue lines in figure \ref{fig:fig_x}, red shading). In fact, because the imposed strain is constant everywhere, the DSP magnitude is approximately the same inside and outside of the frontal region (solid and dashed blue lines in figure \ref{fig:fig_x}). Furthermore, the cancellation between the positive $\langle \text{CP} \rangle_F$ and negative $\langle \text{CP} \rangle_\text{OF}$ values are clearly evident during the \textit{exponential} stage (solid and dashed red lines in figure \ref{fig:fig_x}, blue shading).
The AGSP, which like CP, is determined by the magnitude of the ASC, is considerably more negative when averaged inside the fontal zone (solid and dashed green lines in figure \ref{fig:fig_x}).

The time-integrated $\langle \text{DISP} \rangle$ increases in magnitude from the \textit{exponential} to \textit{superexponential} stages because the NIW wavelength shrinks more rapidly as the front sharpens faster (black line in figure \ref{fig:wKE} and table \ref{tab:table1}). Conversely, the time-integrated $\Delta \langle \mathcal{K} \rangle$ decreases in magnitude from the \textit{exponential} to \textit{superexponential} stages because part of the wave damping due to the AGSP is partially balanced by CP (magenta line in figure \ref{fig:wKE} and table \ref{tab:table1}). Finally, the remaining terms in  (\ref{wave_ke_inviscid}) remove a small amount of  NIW KE during both frontogenetic stages (brown line in figure \ref{fig:wKE} and table \ref{tab:table1}). 
\subsubsection{{The partial cancelation between CP and AGSP during the \textit{superexponential} stage}}
An interesting feature in our solution is that during the \textit{superexponential} stage of frontogenesis, the loss of wave KE due to the AGSP mechanism is partially compensated by the KE gained from ASC via CP (solid red and green lines in the read-shaded region of figure \ref{fig:fig_x} and figures \ref{fig:wKE_superexp}$(d,e)$). 
This can be better understood by projecting the wave momentum flux in the direction of the principal strain axes of the ASC. 
In general, the sum of CP and AGSP in the principal strain coordinates can be expressed as
\begin{align}
\label{sum_cp_agsp}
    \text{CP+AGSP} = \frac{1}{2}({v^\prime}^2+{w^\prime}^2) \bigg(\frac{\partial V}{\partial y}+\frac{ \partial W}{\partial z}\bigg) -({v^\prime}^2-{w^\prime}^2) \frac{S_n^\prime}{2},
\end{align}
where $v^\prime$, $w^\prime$ are velocity components in the transformed coordinates, given by
\begin{subequations}
\label{vel_transform}
\begin{align}
\label{aa}
    v^\prime &= v \cos \theta_p + w \sin \theta_p, \\
\label{bb}
    w^\prime &= - v \sin \theta_p + w \cos \theta_p,
\end{align}
\end{subequations}
and $S_n^\prime$ is 
\begin{align}
    {S_n^\prime}^2={\bigg(\frac{\partial V}{\partial y}-\frac{ \partial W}{\partial z}\bigg)^2+\bigg(\frac{\partial V}{\partial z}+\frac{\partial W}{\partial y}\bigg)^2}.
\end{align}
 The angle between the simulated coordinates and the principal strain coordinates $\theta_p(y,z)$ is given by
\begin{align}
\label{strain_angle}
    \tan 2 \theta_p = \frac{\partial V/\partial z + \partial W
    /\partial y}{\partial V/\partial y - \partial W/\partial z}
    \approx \frac{\partial V/\partial z}{2 \partial V/\partial y},
\end{align}
where the last step is derived from the continuity equation (\ref{eq_gov_e}), assuming $({\partial W}/{\partial y}) \ll ({\partial V}/{\partial z})$, which is valid in our numerical solutions.

{Because our model is $x$-invariant, the first term on right-hand side of (\ref{sum_cp_agsp}) is zero, and thus becomes
\begin{align}
\label{2d_sum_cp_agsp}
    \text{CP+AGSP} = ({w^\prime}^2-{v^\prime}^2) \frac{S_n^\prime}{2}.
\end{align}
When ${w^\prime}^2={v^\prime}^2$ the wave-induced momentum flux in the principal strain coordinates vanish  and CP and AGSP have equal and opposite signs. In this case
$\tan 2 \theta_p \approx -v/(2w)$ and, together with (\ref{strain_angle}), we obtain $w/v=(\partial W/\partial z)/(\partial V/\partial z)$. This implies that the net KE exchanges between NIWs and the ASC are zero only when the phase lines of the NIWs are aligned with the streamlines of the ASC. This particular condition is nearly met during the \textit{superexponential} stage, as the ASC  streamlines align more closely with the isopycnals, and hence with the isophases of the NIWs (figures \ref{fig:strm_exp_suexp}$(a,b)$).
}

\subsubsection{Higher vertical modes}
\label{higher_mode}
The above energetic analysis is solely based on a mode-$1$ NIW interacting with the front. To generalize our results, we initialize the numerical model with mode-$2$ and mode-$3$ Gaussian near-inertial wave packets (\S \ref{sec_niw_mode}) while using the same frontal flow described in \S \ref{semi-anal}, for the case $\alpha=0.1f$. The domain averaged energy exchange terms are computed and summarized in table \ref{tab:table_m2}. 

Qualitatively, the above results for mode-$1$ NIW persist for the higher modes considered here. The $\langle \text{DSP} \rangle$ and $\langle \text{CP} \rangle$ are still the dominant IW energy extraction mechanism during the \textit{exponential} and \textit{superexponential} stages, respectively. Similarly, the $\langle \text{AGSP} \rangle$ causes the wave to lose KE to the ASC during both frontogenetic stages. This suggests that the KE exchange mechanisms are not sensitive to the IW modal structure.

\begin{figure}
    \centering
    \includegraphics[width=\textwidth]{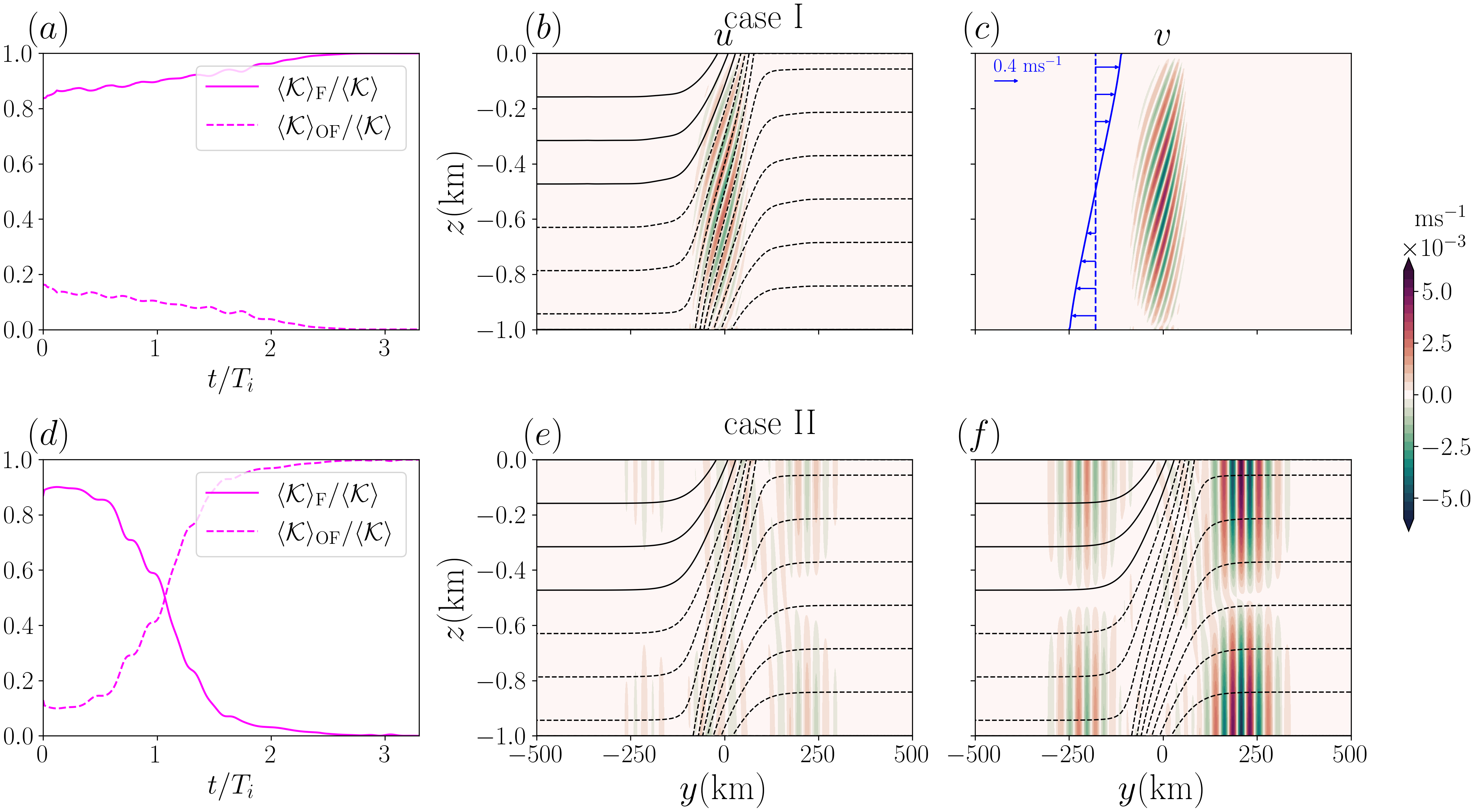}
    \caption{The kinetic energy fractions inside the frontal region ($\langle \mathcal{K}_\text{F}\rangle /\langle \mathcal{K}\rangle$) and outside of it ($\langle \mathcal{K}_\text{OF}\rangle /\langle \mathcal{K}\rangle$) are shown in panels $(a)$ and $(d)$ for minimum frequency (case I) and high-frequency (case II) waves, respectively. The frontal region is identified as the region where $(\partial_y B)^2 > 0.1 (\partial_y B)_{\text{max}}^2$. {Snapshots of mode-$1$ IW velocity components $u$ and $v$ are plotted after two inertial periods in panels $(b,e)$ and $(c,f)$, respectively based on a simulation with $\alpha=0.1f$. The panels $(b,c)$ and $(e,f)$ corresponds to case I and case II, respectively.}
    The black contour lines in figure $(b,e,f)$ display buoyancy $B$ with a $0.016$ ms$^{-2}$ contour interval (solid and dotted line shows positive and negative values, respectively). Blue arrows in panel $(c)$ indicate the profile of the horizontal component of the ASC, $V$, at $y=0$, illustrating that the ageostrophic vertical shear is tilted with the IW phase lines. 
    }
    \label{fig:wave_vel}
\end{figure}
\begin{figure}
\centering
\includegraphics[width=\linewidth]{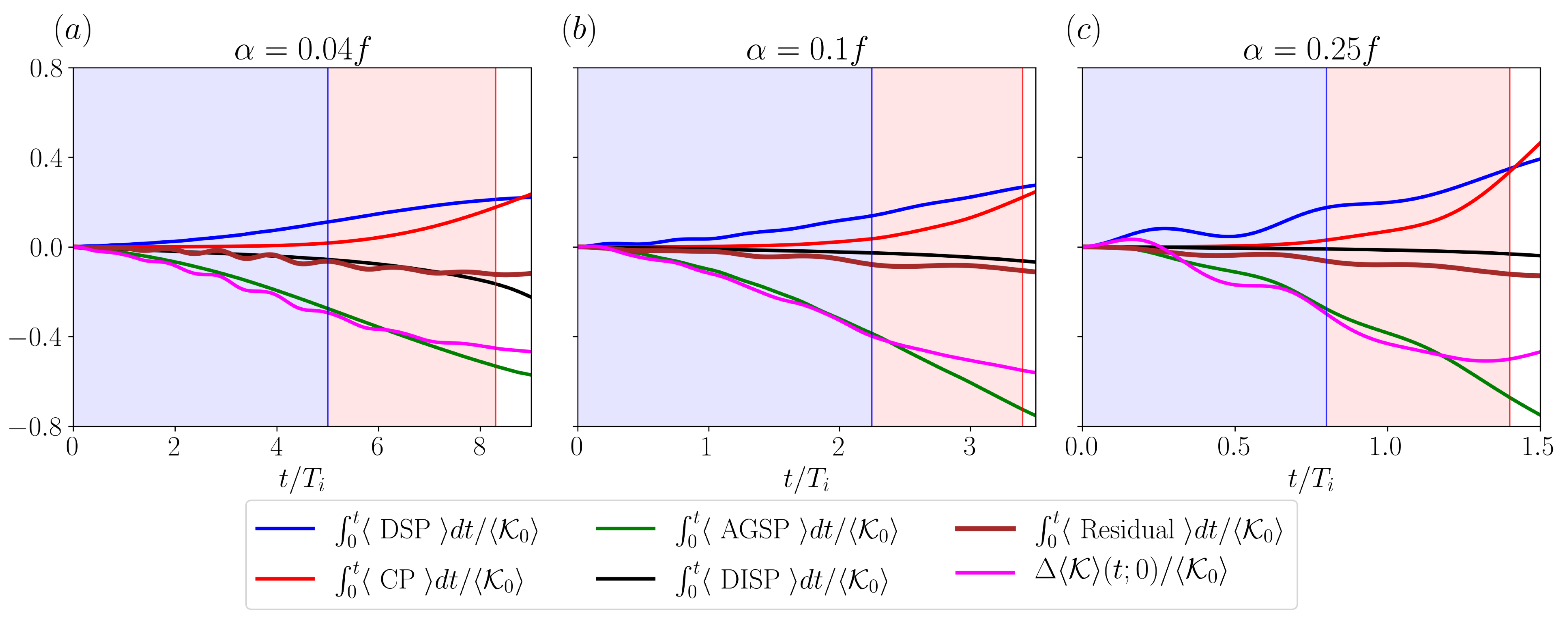}
\caption{The various terms in the IW KE  evolution equation (\ref{wave_ke_inviscid}) for numerical simulations with different $\alpha$ values and a mode-1 IW which is initially near-inertial (Case I). $\text{Residual} = \text{GSP}+\text{LSP}+\text{BFLUX}$ and $\mathcal{K}_0$ is the wave KE at $t_0=0$. 
The end of the \textit{exponential} and \textit{superexponential} frontogenetic stages are denoted by the thin vertical blue and red lines, respectively. Time is normalized by the inertial period $T_i$. 
}
\label{fig:wKE}
\end{figure}
\begin{figure}
    \centering
    \includegraphics[width=\textwidth]{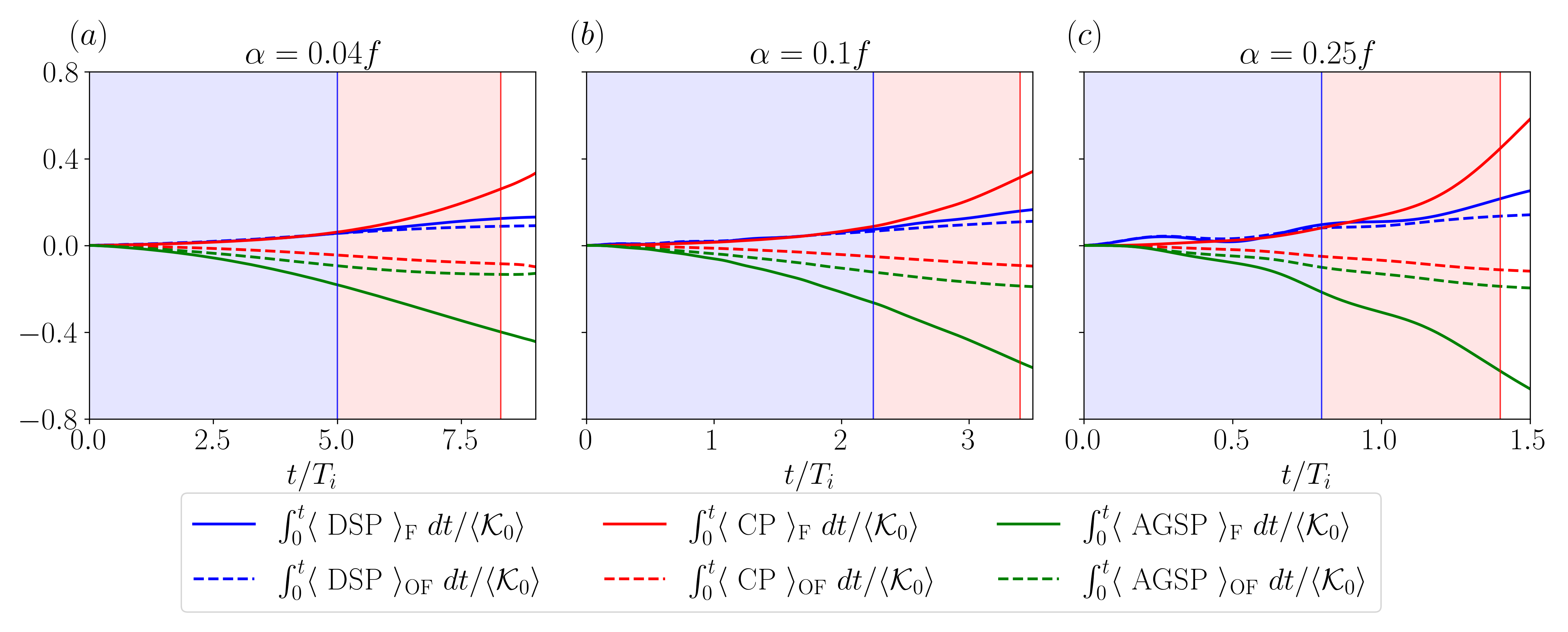}
    \caption{Same as figure \ref{fig:wKE} for the DSP, CP and AGSP terms in the IW KE evolution equation (\ref{wave_ke_inviscid}), averaged over the frontal region defined as the region where $(\partial_y B)^2 > 0.1 (\partial_y B)_{\text{max}}^2$ (subscript F; solid lines) and outside the frontal region (subscript OF; dashed lines).   }
    \label{fig:fig_x}
\end{figure}
\begin{figure}
    \centering
    \includegraphics[width=\textwidth]{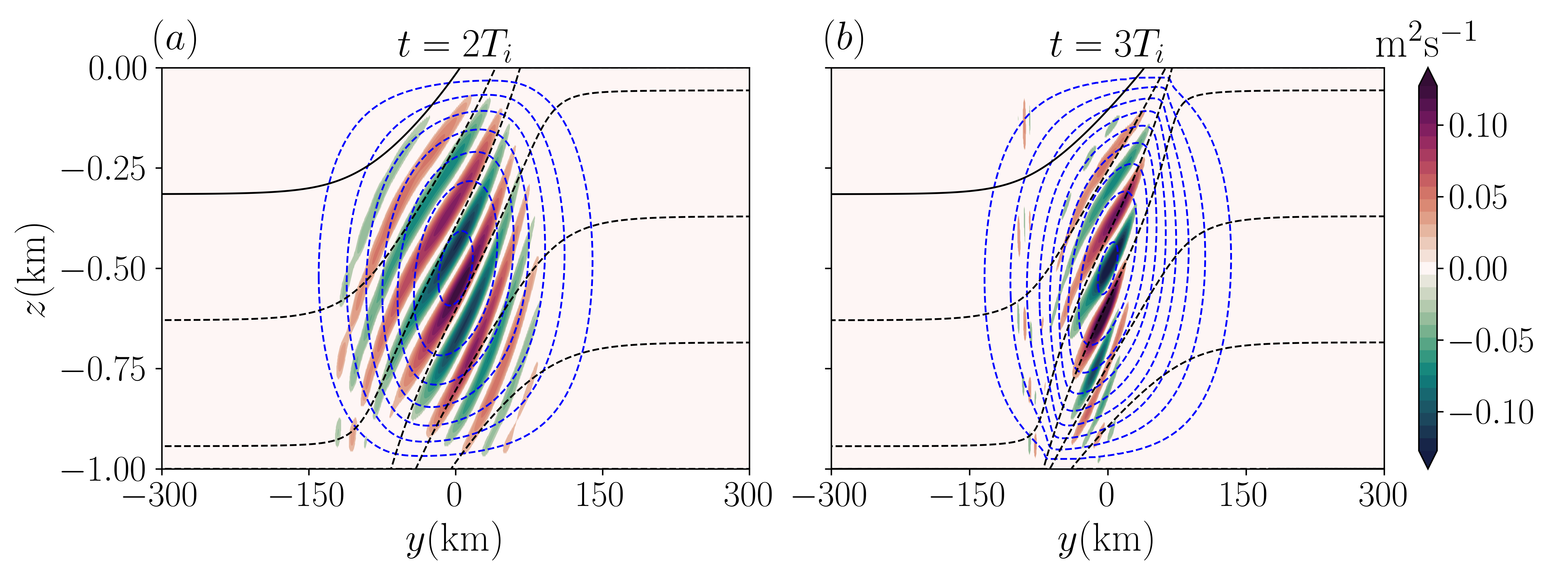}
    \caption{Snapshots of the NIWs streamfunction $\psi$ (color contour plot) with superimposed buoyancy $B$ (black lines) with a $0.03$ ms$^{-2}$ contour interval and ASC streamfunction \textPsi (blue lines) with a $22$ m$^{2}$s$^{-1}$ contour interval during $(a)$ \textit{exponential} ($t=2T_i$, figure \ref{fig:buoy_cmp}($b$)) and $(b)$ \textit{superexponential} ($t=3T_i$) stages of the frontogenesis using numerical simulation for $\alpha=0.1f$. The solid and dotted lines shows the positive and negative values, respectively.}
    \label{fig:strm_exp_suexp}
\end{figure}
\begin{table}\centering
\renewcommand{\arraystretch}{1.9}\setlength\tabcolsep{6pt}
\caption{The various terms in the IW KE evolution equation (\ref{wave_ke_inviscid}) integrated over the exponential and superexponential frontogenetic stages (blue and red shading in figure \ref{fig:buoy_cmp}), for a mode-$1$, minimum frequency IW (Case I) subject to different $\alpha$ values. 
The time integration is from $t_0=0$ to $t=t_e$ for the \textit{exponential} stage, and from $t_0=t_e$ to $t=t_{se}$ for the \textit{superexponential} stage, where $t_e$ and $t_{se}$ denote the end of the \textit{exponential} and the \textit{superexponential} stage, respectively.
}
\label{tab:table1}
\scriptsize
\begin{tabular}{lrrr|rrrr} \toprule
\multirow{2}{*}{Terms} &\multicolumn{3}{c}{\textbf{Exponential}} &\multicolumn{3}{c}{\textbf{Superexponential}} \\\cmidrule{2-7}
&$\alpha=0.04f$ &$\alpha=0.1f$ &$\alpha=0.25f$ &$\alpha=0.04f$ &$\alpha=0.1f$ &$\alpha=0.25f$ \\\cmidrule{1-7}
$\int_{t_0}^t \langle \text{GSP} \rangle dt/\langle \mathcal{K}_0 \rangle$ 
&\multicolumn{1}{c}{-0.03} &\multicolumn{1}{c}{-0.04} &\multicolumn{1}{c}{-0.03} &\multicolumn{1}{c}{0.01} &\multicolumn{1}{c}{-0.03} &\multicolumn{1}{c}{-0.04} \T\B \\
$\int_{t_0}^t \langle \text{LSP} \rangle dt/\langle \mathcal{K}_0 \rangle$ 
&\multicolumn{1}{c}{-0.01} &\multicolumn{1}{c}{-0.01} &\multicolumn{1}{c}{0.00} &\multicolumn{1}{c}{-0.03} &\multicolumn{1}{c}{0.00} &\multicolumn{1}{c}{0.00} \T\B \\
$\int_{t_0}^t \langle \text{DSP} \rangle dt/\langle \mathcal{K}_0 \rangle$ 
&\multicolumn{1}{c}{0.11} &\multicolumn{1}{c}{0.14} &\multicolumn{1}{c}{0.12} &\multicolumn{1}{c}{0.11} &\multicolumn{1}{c}{0.13} &\multicolumn{1}{c}{0.18} \T\B \\
$\int_{t_0}^t \langle \text{CP} \rangle dt/\langle \mathcal{K}_0 \rangle$ 
&\multicolumn{1}{c}{0.02} &\multicolumn{1}{c}{0.04} &\multicolumn{1}{c}{0.04} &\multicolumn{1}{c}{0.16} &\multicolumn{1}{c}{0.18} &\multicolumn{1}{c}{0.31} \T\B \\
$\int_{t_0}^t \langle \text{AGSP} \rangle dt/\langle \mathcal{K}_0 \rangle$
&\multicolumn{1}{c}{-0.28} &\multicolumn{1}{c}{-0.39} &\multicolumn{1}{c}{-0.32} &\multicolumn{1}{c}{-0.25} &\multicolumn{1}{c}{-0.33} &\multicolumn{1}{c}{-0.56} \T\B \\
$\int_{t_0}^t \langle \text{BFLUX} \rangle dt/\langle \mathcal{K}_0 \rangle$ 
&\multicolumn{1}{c}{-0.02} &\multicolumn{1}{c}{-0.03} &\multicolumn{1}{c}{-0.02} &\multicolumn{1}{c}{-0.03} &\multicolumn{1}{c}{-0.03} &\multicolumn{1}{c}{-0.03} \T\B \\
$\int_{t_0}^t \langle \text{DISP} \rangle dt/\langle \mathcal{K}_0 \rangle$
&\multicolumn{1}{c}{-0.06} &\multicolumn{1}{c}{-0.03} &\multicolumn{1}{c}{0.00} &\multicolumn{1}{c}{-0.12} &\multicolumn{1}{c}{-0.04} &\multicolumn{1}{c}{-0.02} \T\B \\
$\Delta \langle \mathcal{K} \rangle (t;t_0)/\langle \mathcal{K}_0 \rangle$
&\multicolumn{1}{c}{-0.27} &\multicolumn{1}{c}{-0.32} &\multicolumn{1}{c}{-0.24} &\multicolumn{1}{c}{-0.15} &\multicolumn{1}{c}{-0.12} &\multicolumn{1}{c}{-0.16} \T\B \\
\bottomrule
\end{tabular}
\end{table}

\begin{table}
\centering
\caption{Same as table \ref{tab:table1} but for mode-$2$ and mode-$3$ NIWs (Case I) and for the $\alpha=0.1f$ solution.
}\label{tab:table_m2}
\scriptsize
\vspace{2mm}
\renewcommand{\arraystretch}{1.9}\setlength\tabcolsep{6pt}
\begin{tabular}{lcc|ccc}\toprule
\multirow{2}{*}{Terms} &\multicolumn{2}{c}{mode-$2$} &\multicolumn{2}{c}{mode-$3$} \\\cmidrule{2-5}
& \textbf{Exponential} & \textbf{Superexponential} 
& \textbf{Exponential} & \textbf{Superexponential} \\\midrule
$\int_{t_0}^t \langle \text{GSP} \rangle dt/\langle \mathcal{K}_0 \rangle$
&\multicolumn{1}{c}{-0.06} &\multicolumn{1}{c}{-0.04} & \multicolumn{1}{c}{-0.05} & \multicolumn{1}{c}{-0.02} \T\B  \\
$\int_{t_0}^t \langle \text{LSP} \rangle dt/\langle \mathcal{K}_0 \rangle$ 
&\multicolumn{1}{c}{0.00} &\multicolumn{1}{c}{0.00} &\multicolumn{1}{c}{0.00} &\multicolumn{1}{c}{0.00} \T\B  \\
$\int_{t_0}^t \langle \text{DSP} \rangle dt/\langle \mathcal{K}_0 \rangle$ 
&\multicolumn{1}{c}{0.12} &\multicolumn{1}{c}{0.07} &\multicolumn{1}{c}{0.09} &\multicolumn{1}{c}{0.06} \T\B \\
$\int_{t_0}^t \langle \text{CP} \rangle dt/\langle \mathcal{K}_0 \rangle$  
&\multicolumn{1}{c}{0.05} &\multicolumn{1}{c}{0.14} &\multicolumn{1}{c}{0.04} &\multicolumn{1}{c}{0.11} \T\B \\
$\int_{t_0}^t \langle \text{AGSP} \rangle dt/\langle \mathcal{K}_0 \rangle$
&\multicolumn{1}{c}{-0.38} &\multicolumn{1}{c}{-0.23} &\multicolumn{1}{c}{-0.31} &\multicolumn{1}{c}{-0.18} \T\B \\
$\int_{t_0}^t \langle \text{BFLUX} \rangle dt/\langle \mathcal{K}_0 \rangle$ 
&\multicolumn{1}{c}{-0.01} &\multicolumn{1}{c}{-0.01} &\multicolumn{1}{c}{0.01} &\multicolumn{1}{c}{-0.02} \T\B \\
$\int_{t_0}^t \langle \text{DISP} \rangle dt/\langle \mathcal{K}_0 \rangle$
&\multicolumn{1}{c}{-0.13} &\multicolumn{1}{c}{-0.11} &\multicolumn{1}{c}{-0.19} &\multicolumn{1}{c}{-0.12} \T\B \\
$\Delta \langle \mathcal{K} \rangle (t;t_0)/\langle \mathcal{K}_0 \rangle$ 
&\multicolumn{1}{c}{-0.41} &\multicolumn{1}{c}{-0.18} &\multicolumn{1}{c}{-0.35} &\multicolumn{1}{c}{-0.17} \T\B \\
\bottomrule
\end{tabular}
\end{table}
%
\subsection{Case II: high-frequency wave}
\label{sec_high_freq}
\begin{figure}
\centering
\includegraphics[width=\linewidth]{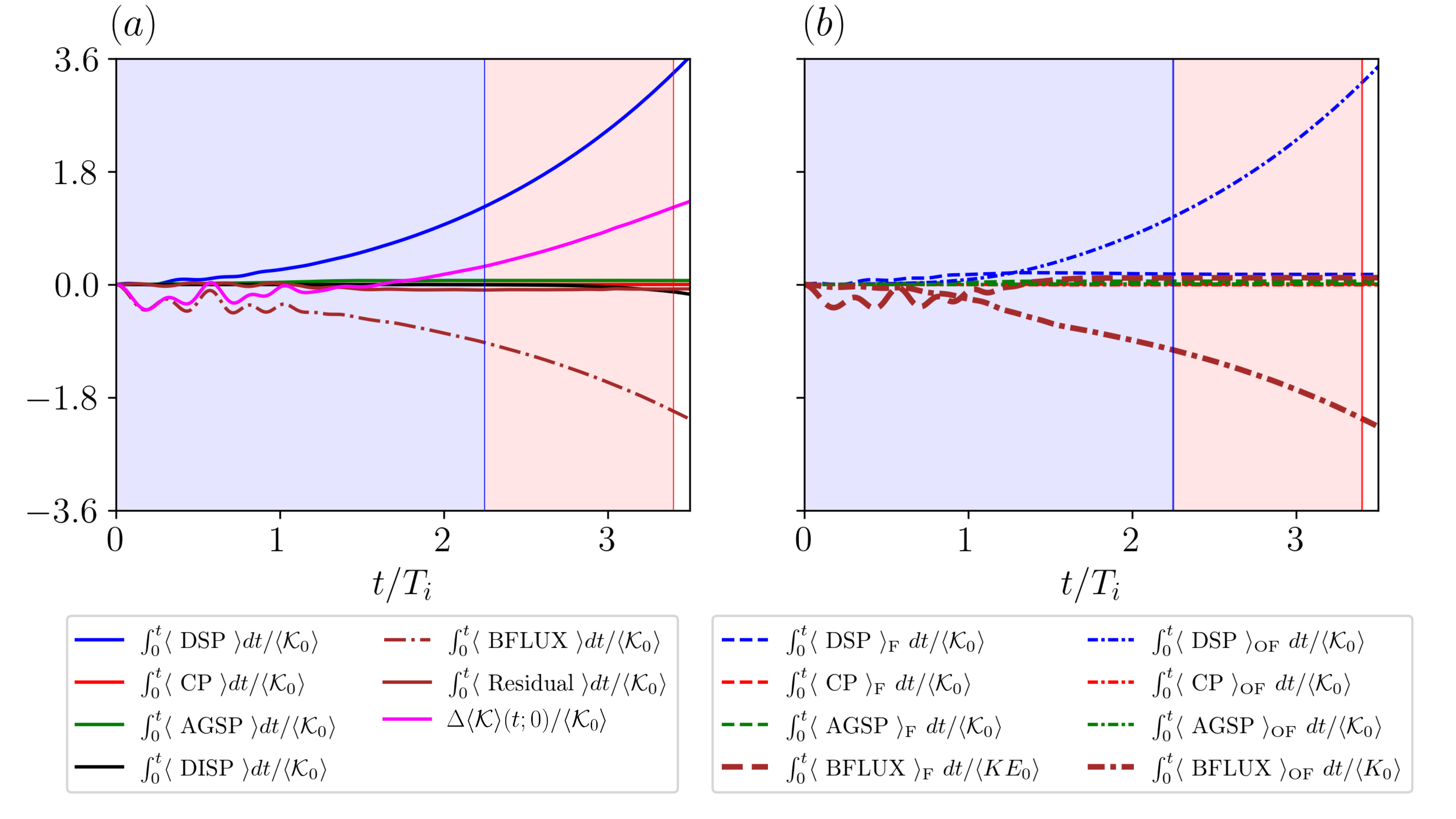}
\caption{$(a)$ Same as figure \ref{fig:wKE}$(b)$ but for a mode-1 high-frequency IW ($\omega=1.5f$; Case II). 
$(b)$ Same as figure \ref{fig:fig_x}$(b)$ but for a mode-1 high-frequency IW ($\omega=1.5f$; Case II). Note that the $\text{Residual} =\text{GSP}+\text{LSP}$.  }
\label{fig:wKE_highf}
\end{figure}

T12 demonstrated that higher-frequency IWs gradually approach the minimum frequency as the front sharpens. In this process, however, the wave phase lines become nearly vertical (figure \ref{fig:wave_vel}$(e,f)$), and the intrinsic horizontal group velocity $c_{g,y} \to -N/m$ (\ref{cg_min_freq}), allowing the wave to escape the frontal region. Due to our configuration setup the IW is unable to propagate out of the imposed-strain influence, as in \cite{thomas2019enhanced}, and is instead halted where $-c_{g,y} = (V-\alpha y)$ (see also supplementary movie 2). Consequently, the KE exchange terms with the front are substantially different than for the NIW (compare figures \ref{fig:wKE}$(b)$ and \ref{fig:wKE_highf}$(a)$, and tables \ref{tab:table1} and \ref{tab:table3}). The IW still gains energy through the $\langle \text{DSP}\rangle$, as the hodographs remain rectilinear (compare figures \ref{fig:wave_vel}$(e)$ and \ref{fig:wave_vel}$(f)$), but this happens outside of the frontal region (blue dot-dashed line in figure \ref{fig:wKE_highf}$(b)$). This is because the strain acts outside of the frontal region, where $Ro \to 0$,  $Ri_g^{-1} \to 0$, and  $|v|/|u| \approx \omega/f > 1$ (\ref{u_eignfun}).
%
The $\langle \text{BFLUX}\rangle$ is now strong and negative (brown lines in figure \ref{fig:wKE_highf}), implying that the wave KE is converted to wave PE (the BFLUX appears with opposite signs in  (\ref{wave_ke_inviscid}) and (\ref{wave_pe_inviscid})). 
{This is consistent with the finding of \cite{xie2015generalised}, where it is shown that the decrease in the horizontal length scale of the wave leads to an increase in wave PE and a subsequent reduction in the Lagrangian-mean balanced kinetic energy. The eulerian-mean energy pathway involves the wave KE equation where through the BFLUX term wave KE is converted to wave PE \citep{rocha2018stimulated}. }
Mechanistically, if the IW phase lines are to remain vertical and steeper than the isopycnals (figure \ref{fig:wave_vel}$(e,f)$), then it must, on average, accumulate PE. 
Because the high-frequency wave remains outside of the frontal region, the AGSP is unable to act and transfer energy back to the front, as is the case for the minimum frequency wave (green line in figure \ref{fig:wKE_highf}), and the high-frequency IW continuously gains KE ($\langle \Delta \text{KE} \rangle >0$ ). The remaining terms are small and are summarized for completeness in table \ref{tab:table3}.

\begin{table}\centering
\renewcommand{\arraystretch}{1.9}\setlength\tabcolsep{6pt}
\caption{Same as table \ref{tab:table1} but for a mode-1 high-frequency IW ($\omega=1.5f$ ; Case II) and for the $\alpha=0.1f$ solution.}
\label{tab:table3}
\scriptsize
\vspace{2mm}
\begin{tabular}{lrrrrr}\toprule
\multirow{2}{*}{Terms} &\multicolumn{2}{c}{\textbf{Exponential phase}} &\multicolumn{2}{c}{\textbf{Superexponential phase}} 
\\\cmidrule{2-5}
$\int_{t_0}^t \langle \text{GSP} \rangle dt/\langle \mathcal{K}_0 \rangle$  &\multicolumn{2}{c}{-0.13} &\multicolumn{2}{c}{-0.05} \\
$\int_{t_0}^t \langle \text{LSP} \rangle dt/\langle \mathcal{K}_0 \rangle$  &\multicolumn{2}{c}{0.05} &\multicolumn{2}{c}{0.02} \\
$\int_{t_0}^t \langle \text{DSP} \rangle dt/\langle \mathcal{K}_0 \rangle$  &\multicolumn{2}{c}{1.23} &\multicolumn{2}{c}{1.92} \\
$\int_{t_0}^t \langle \text{CP} \rangle dt/\langle \mathcal{K}_0 \rangle$  &\multicolumn{2}{c}{0.00} &\multicolumn{2}{c}{0.00} \\
$\int_{t_0}^t \langle \text{AGSP} \rangle dt/\langle \mathcal{K}_0 \rangle$  &\multicolumn{2}{c}{0.06} &\multicolumn{2}{c}{0.00} \\
$\int_{t_0}^t \langle \text{BFLUX} \rangle dt/\langle \mathcal{K}_0 \rangle$  &\multicolumn{2}{c}{-0.93} &\multicolumn{2}{c}{-0.97} \\
$\int_{t_0}^t \langle \text{DISP} \rangle dt/\langle \mathcal{K}_0 \rangle$  &\multicolumn{2}{c}{0.00} &\multicolumn{2}{c}{-0.16} \\
$\Delta \langle \mathcal{K} \rangle(t,t_0)/\langle \mathcal{K}_0 \rangle$ &\multicolumn{2}{c}{0.28} &\multicolumn{2}{c}{0.76} \\
\bottomrule
\end{tabular}
\end{table}

\subsection{Spatial structure of the energy exchange terms}
\label{sec_spatial_features}
\begin{figure}
\centering
\includegraphics[width=\textwidth]{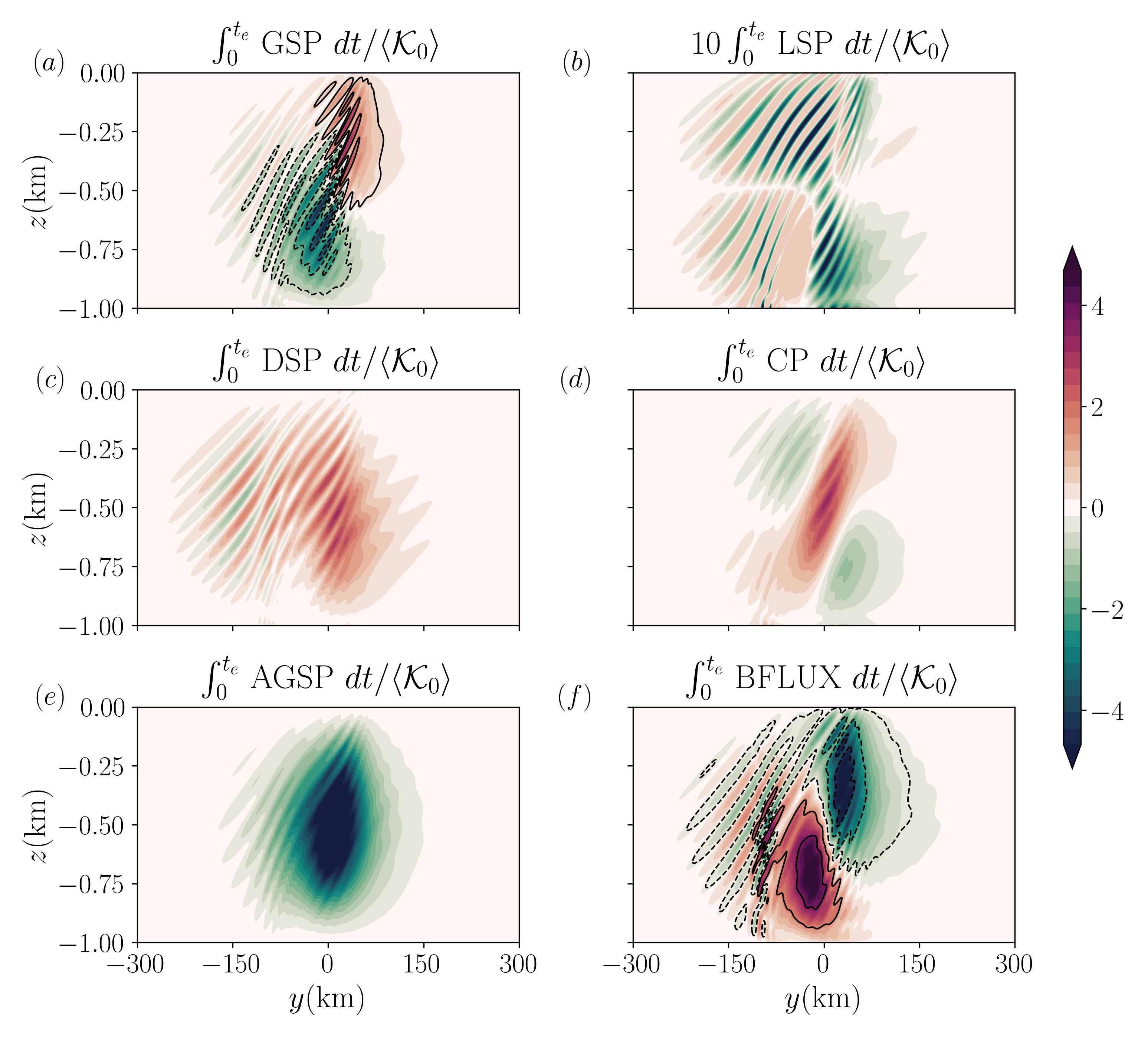}
\caption{Snapshots of the various terms in the IW KE evolution equation (\ref{wave_ke_inviscid}) integrated over the \textit{exponential} frontogenetic stage (blue shading in figure \ref{fig:buoy_cmp}; $t_e$ denotes the end of \textit{exponential} stage), for a mode-$1$, minimum frequency IW (Case I), subject to $\alpha=0.1f$. The approximate GSP and BFLUX values in (\ref{gsp_int}) and (\ref{bflux_int}) are shown with contour-lines in panels $(a)$ and $(f)$, respectively, where solid (dashed) lines denote positive (negative) values with a $2.2$ contour interval for GSP and a $2.4$ contour interval for BFLUX. 
All fields are normalized by the initial, domain-averaged wave KE. 
}
\label{fig:wKE_exp}
\end{figure}

\begin{figure}
\centering
\includegraphics[width=\textwidth]{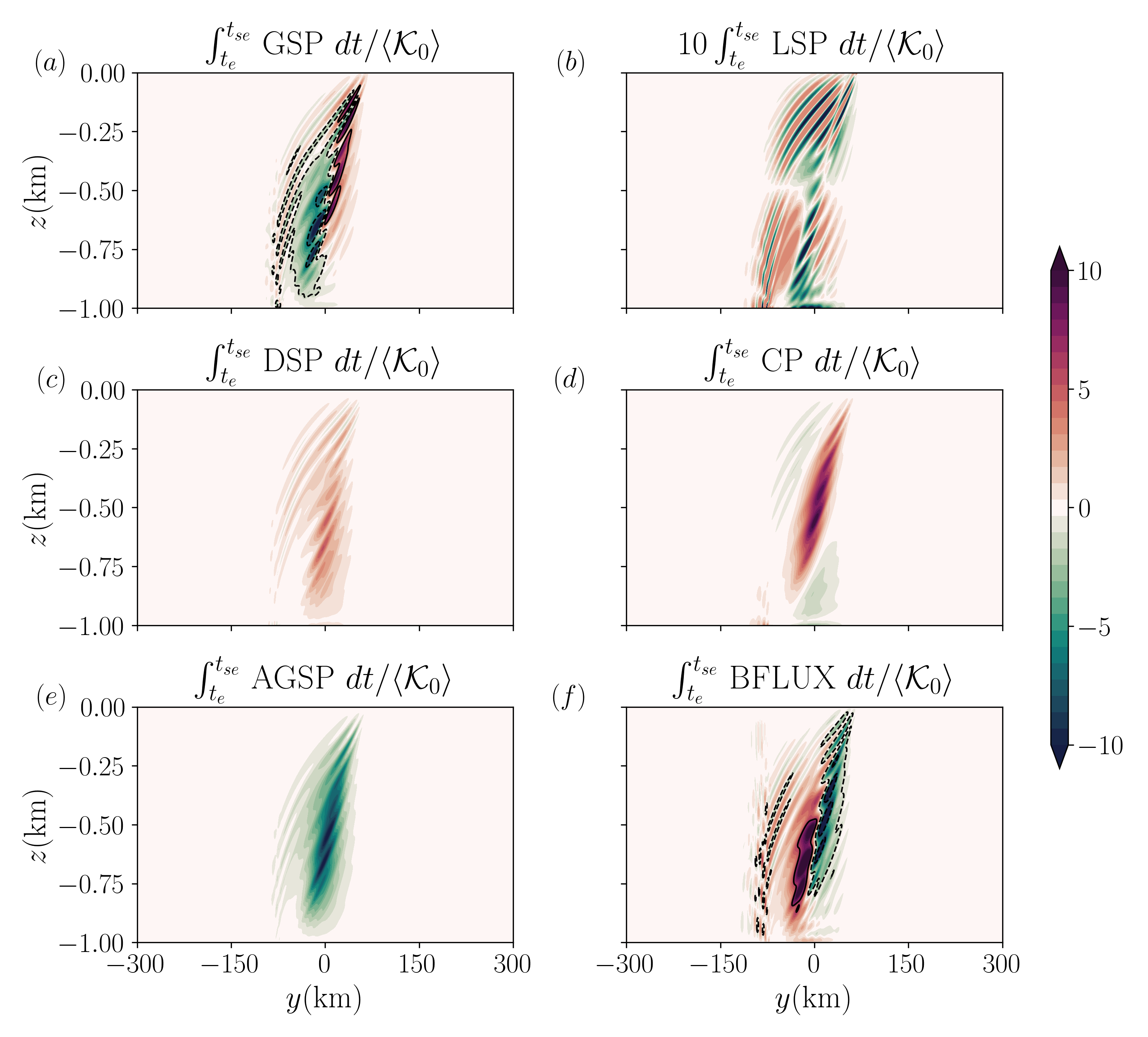}
\caption{Same as figure \ref{fig:wKE_exp}, but time-integrated over the \textit{superexponential} frontogenetic stage (red shading in figure \ref{fig:buoy_cmp}; $t_{se}$ denotes the end of \textit{superexponential} stage).
The contour intervals in panels (a) and (f) are $4.9$ and $6.2$ for the approximate GSP (\ref{gsp_int}) and BFLUX (\ref{bflux_int}), respectively.  
}
\label{fig:wKE_superexp}
\end{figure}

To gain further insight into the IW-front energy exchanges, we examine the spatial structure of the various KE exchange terms in (\ref{wave_ke_inviscid}) during the \textit{exponential} and \textit{superexponential} frontogenetic stages (figures \ref{fig:wKE_exp} and \ref{fig:wKE_superexp}, respectively) for a mode-1 NIW (case I; \S \ref{sec_min_freq}) with $\alpha =0.1f$. 

The time-integrated DSP is predominantly positive during both frontogenetic stages because of the rectilinear wave hodographs (figure \ref{fig:wave_vel}($b,c$)), and is concentrated in the frontal (cyclonic) region (figures \ref{fig:strm_cmp} and \ref{fig:vor_cmp}($a,b$)). As the front sharpens the positive DSP signal is confined to a smaller area with comparable magnitudes in the two frontogenetic stages (figures \ref{fig:wKE_exp}$(c)$ and \ref{fig:wKE_superexp}$(c)$), in agreement with table \ref{tab:table1}. The time-integrated AGSP is negative during both stages because the wave phase lines are tilted with the ageostrophic shear (figure \ref{fig:wave_vel}$(a,b)$), and is even more tightly confined to the frontal region. Similarly to the DSP it occupies a smaller region as the front sharpens, with comparable magnitudes in the two frontogenetic stages (figures \ref{fig:wKE_exp}$(e)$ and \ref{fig:wKE_superexp}$(e)$). 

The cancellation between positive CP in the frontal region and negative CP outside the frontal region during the \textit{exponential} stage (red lines in figure \ref{fig:fig_x}$(b)$) is clearly visible in the spatial plot (figure \ref{fig:wKE_exp}$(d)$). As the convergent ASC strengthens during the \textit{superexponential} stage (figure \ref{fig:wave_vel}$(d)$), CP becomes strongly positive in the frontal region and dominates the negative signal outside the front (figure \ref{fig:wKE_superexp}$(d)$), leading to a domain-averaged positive contribution (figure \ref{fig:wKE}$(b)$ and table \ref{tab:table1}). 

The time-integrated LSP is everywhere an order of magnitude smaller than the remaining terms (figures \ref{fig:wKE_exp}$(b)$ and \ref{fig:wKE_superexp}$(b)$), as expected from table \ref{tab:table1}. 
The time-integrated BFLUX term however (figures \ref{fig:wKE_exp}$(f)$ and \ref{fig:wKE_superexp}$(f)$) exhibits similar magnitudes to the other terms, albeit with both positive and negative lobs that cancel out when averaged over the entire domain (table \ref{tab:table1}). This is because the wave isophases are not exactly parallel to isopycnals but, in fact, have a shallower slope (figures \ref{fig:wave_sol}$(a,b)$). Because the total buoyancy $B+b$ is conserved (as shown below), the wave must acquire a positive (negative) buoyancy anomaly $b$ in the region of lower (higher) $B$. In turn, regions of positive (negative) $b$ are associated with an increase (decrease) in wave PE and consequently, $\text{BFLUX}<0$ ($\text{BFLUX}>0$) (\ref{wave_pe_inviscid}). Interestingly, the integrated GSP term has similar spatial structures to the integrated BFLUX term during both frontogenetic stages, albeit with opposite signs (figures \ref{fig:wKE_exp}$(a,f)$ and \ref{fig:wKE_superexp}$(a,f)$).


To understand this feature in our solutions, we examine the evolution of the total absolute momentum $\mathscr{M}=u+U-fy$ together with the total buoyancy $B+b$, in the inviscid non-diffusive limit. For time scales sufficiently smaller than  $\alpha^{-1}$, it is safe to assume that both the total absolute momentum and total buoyancy are nearly conserved (e.g., T12). Following \cite{whitt2013near}, the $x$-component of the wave velocity $u$ can be written as
\begin{align}
\label{abs_mom}
    u(t+\delta T) - u(t) 
    = -\nabla \mathscr{M}_g \cdot \bm{\delta} \mathbf{r}
    = - \Big( \frac{\partial \mathscr{M}_g}{\partial y} \delta Y
    + \frac{\partial \mathscr{M}_g}{\partial z} \delta Z \Big),
\end{align}
where $\mathscr{M}_g=U - fy$ is the absolute momentum of the geostrophic flow,  $\bm{\delta}\mathbf{r}=\hat{y} \delta Y + \hat{z} \delta Z$ denotes the position vector such that $\delta Y = \int_{t}^{t+\delta T} v dt$ and $\delta Z = \int_{t}^{t+\delta T} w dt$, and $\delta T< \alpha^{-1}$ is the time duration. Similarly, it follows that the wave buoyancy $b$ can be expressed as
\begin{align}
\label{abs_buoy}
    b(t+\delta T) - b(t) = -\nabla B \cdot \bm{\delta} \mathbf{r} 
    = - \Big(\frac{\partial B}{\partial y} \delta Y + \frac{\partial B}{\partial z} \delta Z \Big).
\end{align}
Using (\ref{abs_mom}), the GSP can be written as
\begin{align}
\label{gsp_int}
    -u w \frac{\partial U}{\partial z}\Big\rvert_{t+\delta T} = - \big(u(t) - \nabla \mathscr{M}_g \cdot \bm{\delta} \mathbf{r} \big) w \frac{\partial \mathscr{M}_g}{\partial z}, 
\end{align}
and using (\ref{abs_buoy}), the BFLUX can be expressed as
\begin{align}
\label{bflux_int}
    w b\Big\rvert_{t+\delta T} = w \Big( b(t) - \nabla B \cdot \bm{\delta} \mathbf{r} \Big).
\end{align}
To compute the approximate GSP and BFLUX above we begin with the initial conditions described in (\S \ref{sec_niw_mode}) and time step (\ref{abs_mom}) and (\ref{abs_buoy}) to obtain $u$ and $b$ at all times. 
Equations (\ref{gsp_int}) and (\ref{bflux_int}) are then calculated every $10$ minutes using the numerical values of $v$, $w$, $\mathscr{M}_g$ and $B$.
The approximate GSP and BFLUX fields, time integrated separately over the \textit{exponential} and \textit{superexponential} stages, are shown as contour lines in figures \ref{fig:wKE_exp}$(a,f)$ and \ref{fig:wKE_superexp}$(a,f)$, respectively.
The close resemblance between the approximate and true fields suggests that the spatial structures of the GSP and BFLUX in our solutions are a result of the conservation of total absolute momentum and total buoyancy, respectively.

\section{Summary and Discussion}
\label{summary}
A quasilinear model is developed to study the energy exchanges between a two-dimensional frontal zone undergoing strain-induced semigeostrophic frontogenesis and hydrostatic, linear IW vertical modes. The main novelties of the quasilinear model are:
\smallskip
\begin{enumerate} 
\renewcommand{\theenumi}{\roman{enumi}}
\item $\,$  the frontogenesis includes a \textit{superexponential} sharpening stage that is accompanied by ageostrophic convergent motions and $\Ro \sim \mathcal{O}(1)$ dynamics,
\item  $\,$  the IWs are no longer unbounded in the vertical (e.g., \citealp{thomas2012effects}) and have a modal structure that is more representative of oceanic IWs.
\end{enumerate}
\smallskip
The model is solved numerically for three imposed strain values and IW vertical modes $1-3$ that are initially oriented parallel to isopycnals (case I, minimum frequency NIW) or tilted against isopycnals (case II, high-frequency IWs, $\omega=1.5f$). For all of our solutions we compute the various terms in the wave KE equation (\ref{wave_ke_inviscid}), distinguishing between the \textit{exponential} and \textit{superexponential} frontogenetic stages. 

In agreement with previous work \citep{thomas2019enhanced}, high-frequency waves can escape the frontal zone and, therefore, exchange little energy with the ageostrophic frontal circulation. Nevertheless, because the imposed strain is not only acting in the frontal zone the high-frequency wave can still efficiently extract KE from the balanced deformation flow through the DSP mechanism. Part of this extracted KE is then converted to wave PE.

{NIWs also extract KE from the balanced deformation flow via the deformation shear production (DSP) because the imposed strain modifies the wave hodographs to be rectilinear \citep{thomas2012effects}.
In contrast with high-frequency IWs however, NIW modes remain in the frontal zone and can exchange KE with the ageostrophic frontal flow because their phase-lines align with isopycnal and their group velocity $ c_{g_y} \to 0$.  Indeed, during the exponential phase 
most of the KE extracted due to DSP is transferred to the frontal ASC via the ageostrophic shear production (AGSP), because the wave phase lines are titled with the ageostrophic vertical shear. The inclusion of $\Ro \sim \mathcal{O}(1)$ and ageostrophic convergent motions in our model allows us to identify a new mechanism, the convergence production (CP), through which NIWs can efficiently extract KE from the frontal ageostrophic secondary circulation (ASC).}
In three dimensions (i.e., \citealt{srinivasan2021cascade})
\begin{align}
\label{cp_3d}
\text{CP}\equiv -\delta \mathcal{K},
\end{align}
where $\delta$ denotes the horizontal divergence, and $\mathcal{K}$ is the KE of the IWs. 
{The definition above states that only the sign of  $\delta$ dictates the direction of energy transfer between the wave and mean-flow and that CP plays a role in the energy exchange when the horizontal flow is divergent/convergent. The importance of horizontal divergence to IW energetics has been previously discussed in \cite{weller1982relation} (1982) and \cite{chen2021interaction}, who investigated how a divergent QG/Ekman flow can dampen near-inertial oscillations. 
} Our results demonstrate that CP dominates the energy exchanges in the frontal region during the \textit{superexponential} stage when the convergent ASC inside the frontal zone increases ($|\delta|\sim \mathcal{O}(f)$) and overcomes the divergent flow outside of it. 
{Furthermore, we demonstrate that during the \textit{superexponential} stage the KE loss due to the AGSP mechanism is largely compensated by a KE gain from the ASC due to the CP mechanism. This is because the ASC streamlines are approximately aligned with the isopycnals as well at this stage, and so the NIW momentum fluxes diminish.}

\cite{barkan2021oceanic} demonstrated, using realistically forced high-resolution ocean simulations in the north Atlantic ocean, that the most substantial energy transfers from balanced flow to IWs occur at surface intensified fronts and filaments that are characterized by $\Ro \sim \mathcal{O}(1)$ and strong ageostrophic convergent motions. The results presented here suggest that CP may explain these observed energy transfers. Furthermore, recent numerical results and drifter observations in the Gulf of Mexico indicate that the convergent ASC at oceanic submesoscales (0.1-10 km) may be stronger than predicted by semigeostrophic theory \citep{barkan2019role}, implying that CP may be more significant than is shown by our idealized 2D model. It was further shown in \cite{barkan2019role} that the dynamical dominance of the convergent ASC in oceanic submesoscale fronts and filaments is independent of the physical mechanism that initiated frontogenesis (e.g., turbulent thermal wind; \citealp{gula2014submesoscale}). 
This means that, in contrast with the DSP, CP can lead to energy extraction from oceanic fronts even in the absence of mesoscale straining motions. 

Finally, another important new feature of our quasilinear model is that it can incorporate IW effects on the frontal (`mean') flow by adding quasilinear wave-induced momentum and buoyancy fluxes. These `wave-feedback' effects on frontogenesis, frontal stability, and energy exchanges will be examined in future work.

\section*{Acknowledgements}
SK and RB were supported by Israeli Science Foundation Grant 1736/18.
RB was further supported by NSF Grant OCE-1851397.

\section*{Declaration of Interests}
The authors report no conflicts of interest.

\FloatBarrier
\appendix
\section{Filtering spontaneously emitted internal waves from the mean-flow evolution equations}
\label{filter_igw2}
In this section we describe the procedures to remove spontaneously emitted IWs from the mean-flow equations (\ref{gov_eqs}$a-e$). These emitted IWs are associated with a fast time scale, while the frontogenesis occurs on a slower time scale, provided that $\alpha/f=\epsilon < 1$. To this end we non-dimensionalize (\ref{gov_eqs}$a-e$) and then employ a multiple timescale perturbation approach. The lengths and time are scaled as
\begin{subequations}
\begin{gather}
\label{len_scale}
    y =  \lambda Y^\star, \,\,\,\
    z =  H z^\star, \,\,\,\
    t = \frac{1}{f} t^\star, 
    \tag{\theequation $a-c$}
\end{gather}
\end{subequations}
where $\lambda$ and $H$ are the cross-front and vertical length scales of the front, respectively (Table \ref{tab:params}), and the `star' superscript denotes a non-dimensional variable. The flow variables are scaled with
\begin{subequations}
\begin{gather}
\label{flow_scale}
  (U, V) = |\mathscr{B}| H/(f \lambda) (U^\star, V^\star), \,\,\,\
   W = |\mathscr{B}| H^2/(f \lambda^2) W^\star, \,\,\,\
   P = |\mathscr{B}| H P^\star, \,\,\,\
   B = |\mathscr{B}| B^\star, 
   \tag{\theequation $a-d$}
\end{gather}
\end{subequations} 
where $|\mathscr{B}|$ is the magnitude of the localized front defined in (\ref{eq_b0}). 
With the above non-dimensional variables, the non-dimensional form of (\ref{gov_eqs}$a-e$) is 
\begin{subequations}
\label{evol_non_dim}
\begin{align}
\label{nondim_eq_gov_a}
\frac{DU^\star}{Dt^\star} - V^\star + \epsilon U^\star &= 0, 
\\
\label{nondim_eq_gov_b}
\frac{DV^\star}{Dt^\star} + U^\star - \epsilon V^\star &= -\frac{\partial P^\star}{\partial y^\star},
\\
\label{nondim_eq_gov_c}
0 &= -\frac{\partial P^\star}{\partial z^\star} + B^\star,
\\
\label{nondim_eq_gov_d}
\frac{DB^\star}{Dt^\star} &= 0, 
\\
\label{nondim_eq_gov_e}
\frac{\partial V^\star}{\partial y^\star} + \frac{\partial W^\star}{\partial z^\star} &= 0,
\end{align}
\end{subequations}
where $D/Dt^\star = {\partial}/{\partial t^\star} + (\widetilde{Ro} V^\star - \epsilon y^\star) {\partial}/{\partial y^\star} + \widetilde{Ro} W^\star {\partial}/{\partial z^\star}$, with the Rossby number $\widetilde{Ro}=|\mathscr{B}| H/(f^2 \lambda^2)$ allowed to be $\mathcal{O}(1)$ as in HB72.

Next we decompose all of the mean-flow fields into frontogenetic components (comprising both geostrophic and ageostorphic flows, and denoted by subscript `s'), which evolve over slow time scale $t^\star_s = \epsilon t^\star$, and the spontaneously emitted IW components (denoted by subscript `IW'), which evolve over the fast time scale $t^\star_f = t^\star$, viz. 
\begin{subequations}
\label{decomp_all_non_dim}
\begin{align}
\label{eq_ar11_a}
    U^\star(y^\star, z^\star, t^\star) &= U^\star_s(y^\star, z^\star, t^\star_s) + \epsilon \big(\underbrace{U^\star_s(y^\star, z^\star, t^\star_s)}_{=0} + \eta  U^\star_\text{IW} (y^\star, z^\star, t^\star_f) \big),\\
\label{decomp_2_a}
    B^\star(y^\star, z^\star, t^\star) &= B^\star_s(y^\star, z^\star, t^\star_s) + \epsilon \big(\underbrace{B^\star_s(y^\star, z^\star, t^\star_s)}_{=0}  + \eta B^\star_{\text{IW}} (y^\star, z^\star, t^\star_f) \big), \\
\label{eq_ar21_a}
    P^\star(y^\star, z^\star, t^\star) &= P^\star_s(y^\star, z^\star, t^\star_s) + \epsilon \big(\underbrace{P^\star_s(y^\star, z^\star, t^\star_s)}_{=0}  + \eta P^\star_{\text{IW}} (y^\star, z^\star, t^\star_f) \big), \\
\label{decomp_1_a}
   {V}^\star(y^\star, z^\star, t^\star) &= \underbrace{V^\star_s(y^\star, z^\star, t^\star_s)}_{=0} + \epsilon \big({V}^\star_s(y^\star, z^\star, t^\star_s) + \eta V^\star_{\text{IW}} (y^\star, z^\star, t^\star_f) \big), \\
   \label{decomp_1_aa}
   {W^\star}(y^\star, z^\star, t^\star) &= \epsilon \big({W^\star}_s(y^\star, z^\star, t^\star_s) + \eta W^\star_{\text{IW}} (y^\star, z^\star, t^\star_f) \big),
\end{align}
\end{subequations}
where the $\mathcal{O}(\epsilon^0)$ and $\mathcal{O}(\epsilon^1)$ terms correspond to the geostrophic and ageostrophic frontogenetic flow components, respectively.
$\epsilon=\alpha/f\ll 1$ is a small parameter illustrating that in the theory of semigeostrophic frontogenesis (i.e., HB72 model), the cross-front ASC is always weaker than the along front geostrophic velocity. The variable $\eta=f/S \ll 1$ is another small parameter demonstrating that the magnitude of the spontaneously emitted IWs relative to that of the ASC depends on the strength of the frontal baroclinicity  $S^2=-\partial B/\partial y$. The distinguished limit that allows for a clear ordering separation between frontal and spontaneously-emitted IW fields is $\eta \sim \epsilon^{1/2}$, which is consistent with the parameter regime of our simulations (\S \ref{numerical_model}), and leads to (\ref{decomp_all}$a-e$). Finally, the time derivative is scaled as
\begin{align}
\label{eq_ar31}
    \frac{\partial}{\partial t^\star} = \frac{\partial}{\partial t^\star_f} + \epsilon \frac{\partial}{\partial t^\star_s}.
\end{align}
The evolution of the non-dimensional waveless solutions $(U_s^\star, V_s^\star, W_s^\star, B_s^\star)$ are obtained by substituting (\ref{decomp_all_non_dim}) into (\ref{evol_non_dim}), and truncating the asymptotic series (\ref{decomp_all_non_dim}) at $\mathcal{O}(\epsilon)$.
Equation (\ref{nondim_eq_gov_a}) yields the evolution equation of the waveless $U_s^\star$, which is given by
\begin{align}
\label{evol_u}
    \frac{D U^\star_s}{D t^\star_s} - \epsilon V^\star_s + \epsilon U^\star_s = 0,
\end{align}
where ${D}/{Dt^\star_s}$ is given by 
\begin{align}
\label{nd_eq_gov_flow2}
    \frac{D}{Dt^\star_{s}} = \epsilon \frac{\partial}{\partial t^\star_s} + \epsilon \bigg[ (\widetilde{Ro} V^\star_{s}- y^\star) \frac{\partial}{\partial y^\star} +  \widetilde{Ro} W^\star_{s} \frac{\partial}{\partial z^\star} \bigg].
\end{align}
The following evolution equation for the ageostrophic cross front velocity $V^\star_{s}$ is again obtained by 
truncating (\ref{decomp_all_non_dim}) at $\mathcal{O}(\epsilon)$
\begin{align}
\label{evol_v}
    \epsilon \frac{D V^\star_{s}}{Dt^\star_s} + U^\star_{s} - \epsilon^2 V^\star_{s} = -\frac{\partial P^\star_\text{s}}{\partial y^\star}.
\end{align}
At leading order the above equation yields geostrophic balance for $U_s^\star$.
Combining (\ref{evol_u}) and (\ref{evol_v}) one obtains
\begin{align}
 \frac{D^2 U_s^\star}{Dt_s^\star} + (1 - \epsilon^2) U_{s}^\star = -\frac{\partial P_{s}^\star}{\partial y^\star},
\end{align}
which is the non-dimensional version of (\ref{gov_one}).
From (\ref{nondim_eq_gov_c}) we get hydrostatic balance for $B^\star_\text{s}$ viz.
\begin{align}
\label{nd_eq_gov_flow3}
    0 = -\frac{\partial P^\star_{s}}{\partial z^\star} + B^\star_{s}.
\end{align}
Equation (\ref{nondim_eq_gov_d}) yields the evolution of $B^\star_s$  which is given by
\begin{align}
\label{nd_eq_gov_flow4}
    \frac{DB^\star_{s}}{Dt^\star_s} = 0.
\end{align}
Finally the continuity equation (\ref{nondim_eq_gov_e}) becomes
\begin{align}
\label{nd_eq_gov_flow5}
    \frac{\partial V^\star_{s}}{\partial y^\star} + \frac{\partial W^\star_{s}}{\partial z^\star} = 0.
\end{align}
To summarize, the dimensional form of the evolution equations for the slowly evolving frontogenetic fields are
\begin{subequations}
\label{num_eqs}
\begin{align}
    \frac{D U_{s}}{Dt_s} - fV_\text{s} + \alpha U_{s} &= 0,  \\  
    \frac{D V_\text{s}}{Dt_s} + fU_\text{s} - \alpha V_{s} &= -\frac{\partial P_{s}}{\partial y}, \\
    0 &= -\frac{\partial P_s}{\partial z} + B_s, \\
    \frac{D B_s}{D t_s} &= 0, \\
    \frac{\partial V_s}{\partial y} + \frac{\partial W_s}{\partial z} &= 0,
\end{align}
\end{subequations}
with the material derivative  
\begin{align*}
 \frac{D}{Dt_s}  = \epsilon \frac{\partial}{\partial t_s} + (V_{s}-\alpha y) \frac{\partial}{\partial y} + W_{s} \frac{\partial}{\partial z}.
\end{align*}
Equations (\ref{num_eqs}$a-e$) are solved numerically for the mean-flow variables, as discussed in \S \ref{numerical_model}.


\section{Semi-analytical solution of a uniform PV front}
\label{analytical_sol}
Here we present the semi-analytical solution for a $2$D front undergoing frontogenesis based on the mathematical framework provided by ST13.
Using the definition of the \textit{generalized momentum} coordinates (\ref{gen_coor}$a-c$) and the associated material derivative (\ref{mat_div_gen}),  (\ref{gov_one}) in \textit{generalized momentum} coordinate system becomes
\begin{align}
\label{eq9}
\frac{\partial^2 {U}}{\partial T^2} + \big( f^2 - \alpha^2 \big) {U} = f^2 {U}_g,
\end{align}
where the vertical advection terms are discarded above because, as shown by ST13, their contributions are two-order of magnitude smaller than the linearized solution except when finite-time singularity is reached. 
Substituting the mean-flow buoyancy field (\ref{buoy_field_gmc}) into the PV conservation equation (\ref{311}), and applying the boundary condition 
$(\partial \Delta B/\partial T)=0$ at $Z=-H$ and $Z=0$ 
yields
\begin{align}
\label{def_bprime}
    \Delta B(Y, Z, T) = \frac{N^2}{f} \ee^{\alpha t} 
\int_0^Z \frac{\partial U}{\partial Y} dZ^\prime.
\end{align}
Thus the total buoyancy field ${B}$ can be expressed as
\begin{align}
\label{eq12}
{B}(Y, Z, T) = N^2 Z + B_g(Y) + \frac{N^2}{f} \ee^{\alpha T} \int_0^Z \frac{\partial {U}}{\partial Y} dZ^\prime.
\end{align}
Combining (\ref{eq_gov_d}) and (\ref{eq12}) we obtain an expression for the vertical velocity $W$ 
\begin{align}
\label{eq_315}
W = -\frac{\mathcal{J}}{f} \ee^{-\alpha T} \int_0^Z \frac{\partial}{\partial T} \Big(\ee^{\alpha T} \frac{\partial {U}}{\partial Y} \Big) dZ^\prime,
\end{align}
where the Jacobian $\mathcal{J}$ is defined in (\ref{def_J}).
It is convenient to introduce a cross-front streamfunction {\textPsi}$(Y,Z)$
\begin{align}
\label{eq_psi}
\text{\textPsi} = - \int W dy = - \int_{-\infty}^Y {W} \mathcal{J}^{-1} dY = \frac{1}{f} \ee^{-\alpha T}  \frac{\partial}{\partial T} \Big( \ee^{\alpha T} \int_0^Z dZ^\prime {U} \Big),
\end{align}
where for evaluating the $Y$ integral of the above equation we assume that ${U} \to 0$ at $Y \to \pm\infty$. 
The associated cross-front velocity ${V}$ can be expressed as 
\begin{align}
\label{eq_120}
{V} 
= \frac{1}{f} \Bigg[ \Big( \frac{\partial {U}}{\partial T} + \alpha {U} \Big) +  {W} \frac{\partial {U}}{\partial Z} \Bigg].
\end{align}
The thermal wind relation (\ref{eq_14}) becomes
\begin{align}
\label{tw}
f \ee^{-\alpha T} \frac{\partial {U}_g}{\partial Z} + 
\frac{\partial {B}}{\partial Y} = \frac{\partial {U}}{\partial Z} \frac{\partial {U}_g}{\partial Y} - 
\frac{\partial {U}_g}{\partial Y}\frac{\partial U}{\partial Y},
\end{align}
where the right-hand side of the above equation constitutes a Jacobian, which is zero when the along-front velocity $U$ is a functional form of the along-front geostrophic velocity $U_g$. In the HB72 model ${U}={U}_g$, so the right-hand side of (\ref{tw}) is identically zero, such that
\begin{align}
\label{eq_ug}
{U} = -\frac{1}{f} \ee^{\alpha T} \int 
\frac{\partial {B}}{\partial Y} dZ.
\end{align}
With the above definition of ${U}$, (\ref{eq9}) becomes  (\ref{eq_123}), which can be solved by assuming that ${U}$ takes the following form
\begin{align}
\label{eq_124}
{U}(Y, Z, T) = \sum_{n=1}^\infty \cos (\mathfrak{m}_n Z) \int_{-\infty}^{\infty} \widehat{U}(\mathfrak{l}, \mathfrak{m}_n, T) \ee^{\ii \mathfrak{l} Y} d\mathfrak{l},
\end{align}
where $\mathfrak{l}$ and $\mathfrak{m}_n=n\pi/H$ are the horizontal and vertical wavenumbers, respectively, and `hat' denotes the Fourier mode amplitude. 
{Note that we choose cosine modes for $U$ in the vertical direction to satisfy the free-slip boundary conditions.}
Substituting (\ref{eq_124}) into (\ref{eq_123}) we obtain
\begin{align}
\label{eq_125}
\frac{\partial^2 \widehat{U}}{\partial T^2} + \Bigg[ (f^2 - \alpha^2)  + N^2 \ee^{2\alpha T} \frac{\mathfrak{l}^2}{\mathfrak{m}_n^2} \Bigg] \widehat{U} = -\ii f \mathfrak{l} \ee^{\alpha T}  
\widehat{B}_g \mathcal{A}_n,
\end{align}
with
\begin{align}
\mathcal{A}_n 
= -\frac{2H}{n^2 \pi^2} \big[-1 + (-1)^n \big],
\end{align}
and subject to the assumption that the right-hand-side of (\ref{eq_123}) vanishes because the frontogenetic flow is purely baroclinic. 
{
Defining {\textPsi} and $\Delta B$ similarly to (\ref{eq_124}), where both involve sine modes in the $z$-direction to satisfy the no penetration and zero buoyancy perturbation respectively}, and making use of (\ref{eq_psi}) and (\ref{def_bprime}), we obtain 
\begin{subequations}
\begin{align}
\label{eq_331}
\widehat{\text{\textPsi}} &= \frac{1}{f} \frac{1}{\mathfrak{m}_n} \Big( \frac{\partial \widehat{U}}{\partial T} + \alpha \widehat{U} \Big),
\\
\label{eq_333}
\widehat{\Delta B} &= \ii \frac{N^2}{f} \frac{\mathfrak{l}}{\mathfrak{m}_n} \ee^{\alpha T} \widehat{U}.
\end{align}
\end{subequations}
The general solution of (\ref{eq_125}) consists of two parts. First, a homogeneous part that is associated with spontaneously emitted IWs and, second, an inhomogeneous part that is associated with strain-induced frontogenesis. 
Because the focus of this study is on frontogenesis, we modify (\ref{eq_125}) to obtain a waveless frontogenetic solution, as is outlined in the next section.

\subsection{Filtering spontaneously emitted high-frequency IWs from the solution}
\label{filter_igw1}
\begin{figure}
    \centering
    \includegraphics[width=\textwidth]{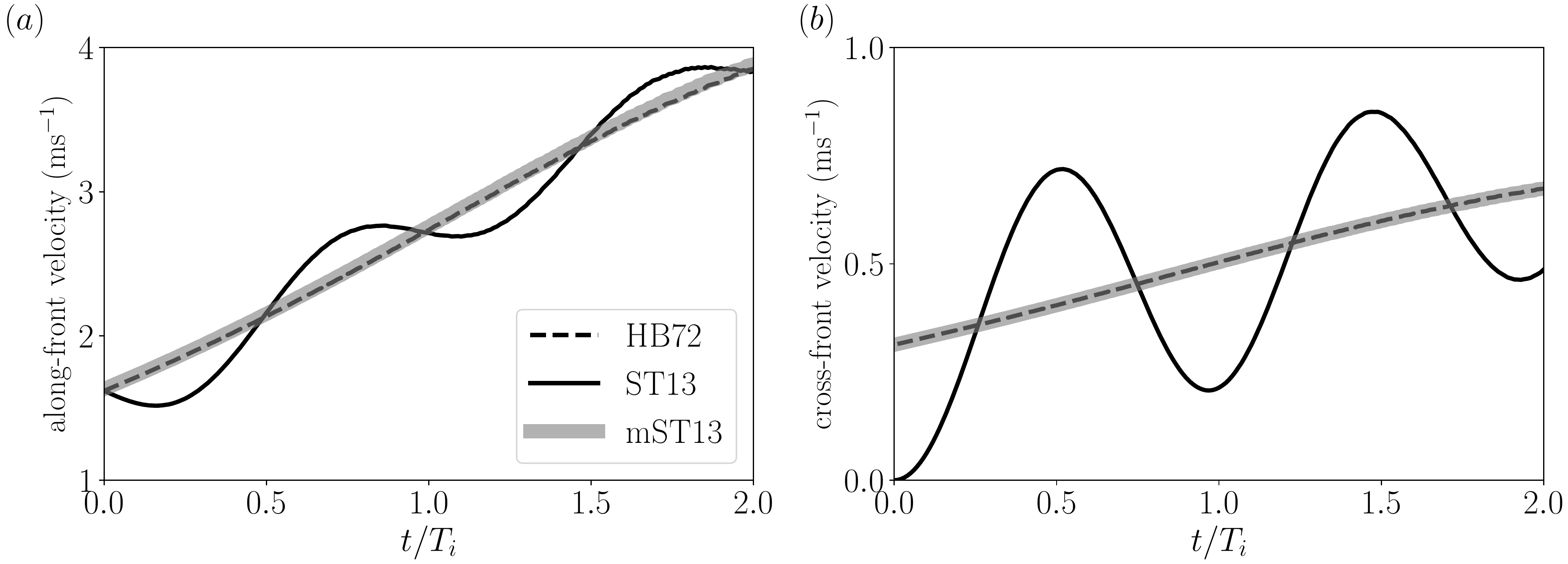}
    \caption{The evolution of (a) the along-front velocity $U(y,z,t)$ and (b) the cross-front velocity $V(y,z,t)$ at the location of the maximum horizontal buoyancy gradient for the case of $\alpha=0.1f$. 
    HB72 is the analytical solution of HB72, ST13 is the analytical solution of ST13, and mST13 is the `modified' ST13 solution described in (\ref{mst13}, \ref{eq_app331}). Time is normalized by the inertial period $T_i$. 
     }
    \label{fig:sol_cmp}
\end{figure}

Following the same methodology discussed in Appendix \ref{filter_igw2} and making use of the characteristics length and time scales defined in (\ref{len_scale}), 
(\ref{eq_123}) non-dimensionalizes to
\begin{align}
\label{nondim_analytic}
    \frac{\partial^2 U^\star}{{\partial T^\star}^2} + (1 - \epsilon^2) U^\star + Bu \ee^{2 \epsilon T^\star} \int \int_0^{Z^\star} \frac{\partial^2 U^\star}{{\partial Y^\star}^2} d{Z^\star}^\prime dZ^\star = -
    \ee^{\epsilon T^\star} \frac{dB_g^\star}{dY^\star} \int dZ^\star,
\end{align}
where the Burger number
\begin{align}
\label{burger_num}
    {Bu} = \frac{N^2 H^2}{f^2 \lambda^2}.
\end{align}
Next, we apply the same decomposition as in (\ref{eq_ar11_a}), using the distinguished limit $\eta\sim \epsilon^{1/2}$, 
\begin{align}
\label{eq_ar1}
    U^\star(Y^\star, Z^\star, T^\star) = U^\star_\text{s}(Y^\star, Z^\star, T^\star_s) + \epsilon^{\tfrac{3}{2}}  U^\star_{\text{IW}} (Y^\star, Z^\star, T^\star_f), 
\end{align}
where $T^\star_f = T^\star$ denotes the fast time scale and
$T^\star_s = \epsilon T^\star$ denotes the slow time scale.
The associate time derivative scales as
\begin{align}
\label{eq_ar2}
    \frac{\partial}{\partial T^\star} = \frac{\partial}{\partial T^\star_f} + \epsilon \frac{\partial}{\partial T^\star_s}.
\end{align}
Substituting (\ref{eq_ar1}) and (\ref{eq_ar2}) into (\ref{nondim_analytic}) and truncating the asymptotic series (\ref{eq_ar11_a}) at $\mathcal{O}(\epsilon)$ yields
\begin{align}
\label{evol_U}
    \epsilon^2 \frac{\partial^2 \widehat{U^\star_\text{s}}}{{\partial T^\star_s}^2} 
     + \Big[ 1 - \epsilon^2 + Bu \ee^{2 \epsilon T^\star} \frac{{\mathfrak{l}^\star}^2}{{\mathfrak{m}^\star}_n^2} \Big] \widehat{U^\star_\text{s}} = - \ii \mathfrak{l}^\star \ee^{\epsilon T^\star} \widehat{B^\star_g} \mathcal{A}^\star_n,
\end{align}
where the $\mathfrak{l}^\star$ and $\mathfrak{m}^\star_n$ are the non-dimensional horizontal and vertical wavenumbers, respectively, and $\mathcal{A}_n^\star = -{2}/{(n^2 \pi^2)} \big[-1 + (-1)^n \big]$.
Equation (\ref{evol_U}) is transformed back to the following dimensional form 
\begin{align}
\label{mst13}
    \epsilon^2 \frac{\partial^2 \widehat{U_\text{s}}}{{\partial T_s}^2} + \Big[f^2 - \alpha^2 + N^2 \ee^{2 \alpha T_f} \frac{\mathfrak{l}^2}{\mathfrak{m}_n^2} \Big] \widehat{U_\text{s}} &= - \ii f \mathfrak{l} \ee^{\alpha T_f} \widehat{B_g} \mathcal{A}_n,
\end{align}
which is valid for timescales of $\mathcal{O}(\alpha^{-1})$.
We refer to the above equation as `modified' ST13. At $\mathcal{O}(1)$ of (\ref{mst13}) yields the HB72 solution which is given by
\begin{align}
\label{hb72}
    \widehat{U_\text{s}} = -\frac{\ii f \mathfrak{l} \ee^{\alpha T_f} \widehat{B_g} \mathcal{A}_n}{f^2 + N^2 \ee^{2 \alpha T_f} \frac{\mathfrak{l}^2}{\mathfrak{m}_n^2}}
\end{align}
The ASC $\widehat{\text{\textPsi}_\text{s}}$ and buoyancy deviation $\widehat{\Delta B_\text{s}}$ are obtained using (\ref{eq_331}) and (\ref{eq_333})
\begin{subequations}
\begin{align}
\label{eq_app331}
\widehat{\text{\textPsi}_\text{s}} &= \frac{1}{f} \frac{1}{\mathfrak{m}_n} \Big( \frac{\partial \widehat{U_\text{s}}}{\partial T_s} + \alpha \widehat{U_\text{s}} \Big),
\\
\label{eq_app333}
\widehat{\Delta B_\text{s}} &= \ii \frac{N^2}{f} \frac{\mathfrak{l}}{\mathfrak{m}_n} \ee^{\alpha T_f} \widehat{U_\text{s}}.
\end{align}
\end{subequations}
The along-front and cross-front velocities for the HB72, ST13, and `modified' ST13 analytical solutions are computed at the location of the maximum horizontal buoyancy gradient for the case of $\alpha=0.1f$ (figure \ref{fig:sol_cmp}$(a,b)$).
The ST13 solution oscillates about the HB72 solution because it contains spontaneously emitted waves, while the waveless `modified' ST13 solution closely resembles the HB72 solution. 

\section{Geostrophic adjustment of the initial state}
\label{analytic_ic}
ST13 demonstrated that the initial buoyancy field $B$ (\ref{buoy_field_gmc}) and the associated along-front geostrophic velocity $U$ (\ref{eq_ug}) have an unbalanced part that adjusts to a geostrophic steady-state, emitting IWs in the process.
Here we follow the same procedure outlined in ST13 to obtain a steady-state initial condition and suppress IW emission due to geostrophic adjustment. In the absence of any imposed strain ($\alpha=0$), (\ref{eq_125}) yields
\begin{align}
\label{eq_127}
  \frac{\partial^2 \widehat{U}}{\partial T^2} + \omega_{\text{IW}}^2 \widehat{U} = -\ii f  l \widehat{B}_g \mathcal{A}_n,
\end{align}
where the hydrostatic IW frequency
\begin{align}
\label{eq_128}
    \omega_{\text{IW}} = f \sqrt{1  +  \frac{N^2 l^2}{f^2 \mathfrak{m}_n^2}}.
\end{align}
Following \cite{blumen2000inertial}, 
we set the initial condition to zero motion i.e., ${U}={V}={W}=0$.
From (\ref{eq_120}) we further get $\partial {U}/\partial T = 0$. With these initial conditions the solution to (\ref{eq_127}) is
\begin{align}
\label{eq_129}
    \widehat{U} = -\ii \frac{l f \widehat{B}_g \mathcal{A}_n}{\omega_{\text{IW}}^2} \big[1 - \cos({\omega_{\text{IW}} T}) \big].    
\end{align}
The wave solutions for {\textPsi}  and $\Delta B$ are obtained by substituting (\ref{eq_129}) into (\ref{eq_331}) and (\ref{eq_333}), respectively,
\begin{subequations}
\begin{align}
\label{eq_130}
\widehat{\text{\textPsi}} &= -\ii \frac{l \widehat{B}_g\mathcal{A}_n}{\mathfrak{m}_n \omega_{\text{IW}}} \sin({\omega_{\text{IW}} T}), 
\\
\label{eq_131}
\widehat{\Delta B} &= \frac{l^2 \widehat{B}_g \mathcal{A}_n N^2}{\mathfrak{m}_n \omega_{\text{IW}}^2} \big[1 - \cos({\omega_{\text{IW}} T}) \big].     
\end{align}
\end{subequations}
The geostrophically adjusted solutions, which are used to generate waveless initial conditions, are obtained by taking only the time independent part of (\ref{eq_129}), (\ref{eq_130}), and (\ref{eq_131})
\begin{subequations}
\label{final_state}
\begin{align}
\label{ss_1}
    \widehat{U}_0 &= -\ii \frac{l f \widehat{B}_g \mathcal{A}_n}{\omega_{\text{IW}}^2}, \\
\label{ss_2}
    \widehat{\text{\textPsi}}_0 &= 0, \\
\label{ss_3}
    \widehat{\Delta B_0} &= \frac{l^2 \widehat{B}_g\mathcal{A}_n N^2}{\mathfrak{m}_n \omega_{\text{IW}}^2}.
\end{align}
\end{subequations}

\begin{figure}
    \centering
    \includegraphics[width=\textwidth]{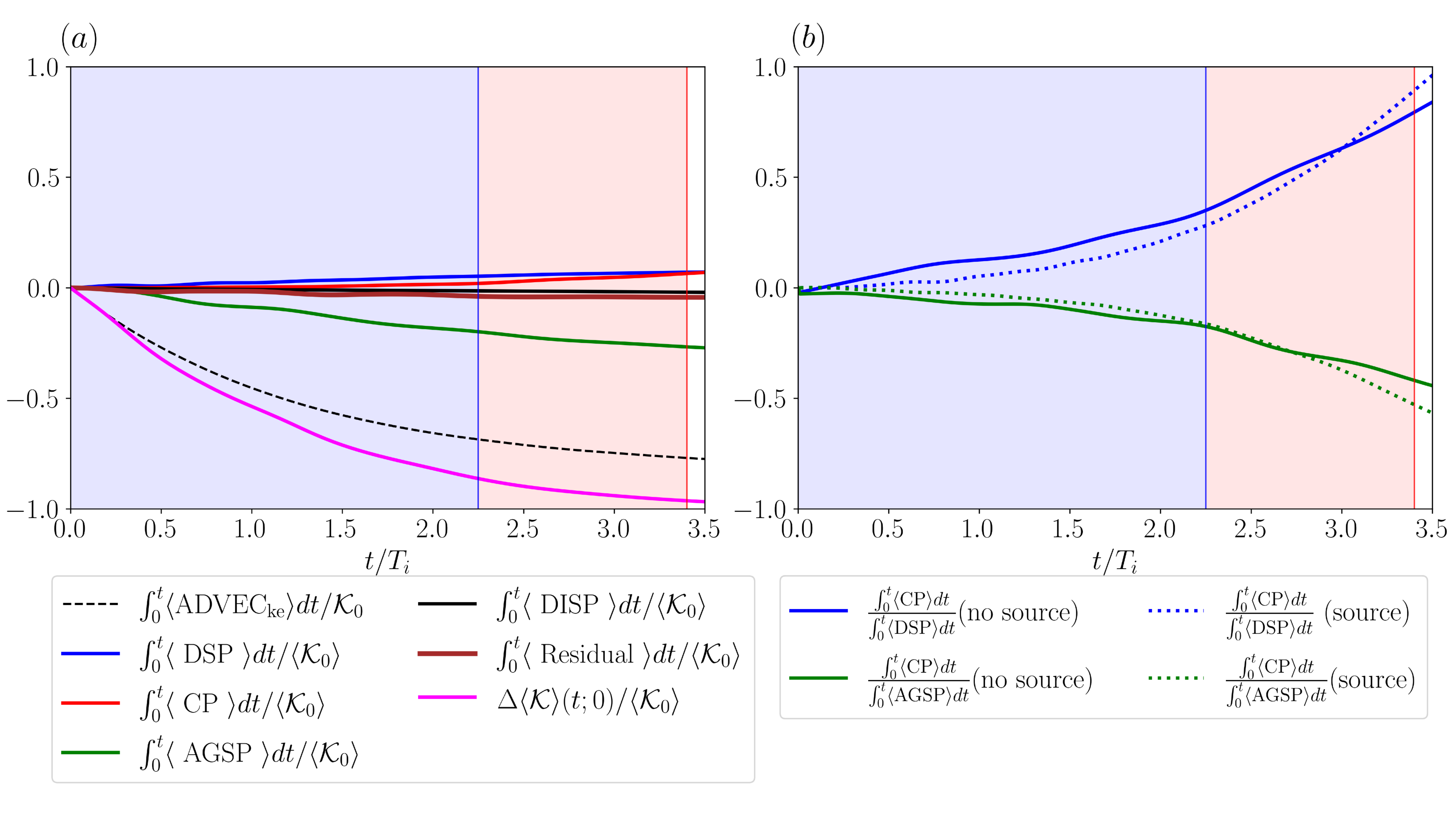}
    \caption{$(a)$ Same as figure \ref{fig:wKE}$(b)$ but without the source terms ($\mathcal{F}_u$, $\mathcal{F}_v$ and $\mathcal{F}_b$) in the wave equations (\ref{eq_wave_all_1a}$a-e$). The term $\text{ADVEC}_\text{ke}$ is given by $\text{ADVEC}_\text{ke}= (V-\alpha y) \partial \mathcal{K}/\partial y + W \partial \mathcal{K}/\partial y$ in (\ref{wave_ke_inviscid}).
    $(b)$ The ratio between time-integrated, and domain averagedx DSP and CP (blue lines) and AGSP and CP (green lines) for the two cases - without (solid lines) and with (dotted lines) the source terms in the wave evolution equations. The results plotted here are for $\alpha=0.1f$. The results for the other strain values discussed in the manuscript are summarized in table \ref{tab:table1a}.
    }
    \label{fig:comparison}
\end{figure}

The corresponding geostophically adjusted initial condition of the buoyancy field (\ref{buoy_field_gmc}) in the \textit{generalized momentum} coordinate system is thus
\begin{align}
\label{buoy_ss}
    B_{0}(Y, Z) = N^2 Z + B_g(Y) + \Delta B_\text{0}(Y, Z), 
\end{align}
where $\Delta B_{0}$ is obtained from (\ref{ss_3}). Similarly, the geostrophically adjusted initial along-front velocity $U_{0}$ is obtained from (\ref{ss_1}) and $V_0=W_0=0$ (\ref{ss_2}). 
The semi-analytical solutions describe in the manuscript are obtained by integrating 
(\ref{mst13}) numerically, subject to the initial conditions 
\begin{subequations}
\begin{gather}
\widehat{U}_\text{s}(T=0) = \widehat{U}_{0}, 
    \,\,\,\,\,\,\
\frac{\partial \widehat{U}_\text{s}(T=0)}{\partial T} = -\alpha  \widehat{U}_{0},
\tag{\theequation $a-b$}
\end{gather}
\end{subequations}
where the last condition is obtained from (\ref{eq_331}) and (\ref{ss_2}). 
Time-stepping is performed using a $4$th-order Runge-Kutta scheme with a time-step of $20$ s, a domain size $Y \in [-1000, 1000]$ km and $Z \in [-1, 0]$ km, and with $1500$ Fourier modes in the $Y$-direction and $240$ cosine modes in the $Z$-direction. 
{\section{Quantifying the effects of $\mathcal{F}_u$, $\mathcal{F}_v$ and $\mathcal{F}_b$ in equations (\ref{eq_wave_all}$a-e$)}}
\label{wo_forcing}
As discussed in \S \ref{numerical_model} the terms $(\mathcal{F}_u, \mathcal{F}_v, \mathcal{F}_b) \equiv 1/2(\alpha u, \alpha v, \alpha b)$ are added to the IW momentum and buoyancy equations to ensure energy conservation. It can be verified the these terms compensate for the energy sink due to the imposed geostrophic strain, which is horizontally divergent in the $x$-invariant numerical configuration we use. To demonstrate that these terms do not affect the physics associated with the energy exchange mechanism discussed in the manuscript we re-run our numerical simulations without these source terms for the same values of $\alpha$ and for a mode-$1$ minimum frequency IW. As expected, a significant amount of wave KE drains out of the domain due to the advection induced by the imposed strain (dotted line in figure \ref{fig:comparison}$a$) and, consequently, the absolute magnitudes of the most significant energy exchange terms are reduced (figure \ref{fig:comparison}$a$). Nonetheless, the ratios between the time-integrated and domain-averaged CP and DSP, and CP and AGSP (figure \ref{fig:comparison}$(b)$ and table \ref{tab:table1a}) are essentially unaffected, and are consistent with the results discussed in \S \ref{sec_min_freq}. This shows that these terms have negligible impact on the energy exchange processes discussed in the manuscript.

\begin{table}
\centering
\renewcommand{\arraystretch}{1.9}\setlength\tabcolsep{6pt}
\caption{{A summary of the domain-averaged energy exchange terms for solutions with and without the source terms in the wave evolution equation (\ref{eq_wave_all}), integrated separately over the exponential and superexponential frontogenetic stages (blue and red shading in figure \ref{fig:buoy_cmp}), for a mode-$1$, minimum frequency IW (Case I). Residual=GSP+LSP+BFLUX. The time integration is from $t_0=0$ to $t=t_e$ for \textit{exponential}, and from $t_0=t_e$ to $t=t_{se}$ for \textit{superexponential} stage. The variables $t_e$ and $t_{se}$ denote the end of \textit{exponential} and \textit{superexponential} stage, respectively.
} }
\label{tab:table1a}
\scriptsize
\begin{tabular}{lrrrr|rrrr}\toprule
\multirow{2}{*}{Sources} &\multirow{2}{*}{Ratios} &\multicolumn{3}{c}{\textbf{Exponential}} &\multicolumn{3}{c}{\textbf{Superexponential}} \\\cmidrule{3-8}
& &$\alpha=0.04f$ &$\alpha=0.1f$ &$\alpha=0.25f$ &$\alpha=0.04f$ &$\alpha=0.1f$ &$\alpha=0.25f$ 
\\\midrule
\multirow{4}{*}{Yes} 
&$\dfrac{\int_{t_0}^{t} \langle \text{CP} \rangle dt}{\int_{t_0}^{t} \langle \text{DSP} \rangle dt}$
&\multicolumn{1}{c}{0.18} &\multicolumn{1}{c}{0.30} &\multicolumn{1}{c}{0.33} &\multicolumn{1}{c}{1.45} &\multicolumn{1}{c}{1.38} &\multicolumn{1}{c}{1.72} \T\B \\[10pt]
&$\dfrac{\int_{t_0}^{t} \langle \text{CP} \rangle dt}{\int_{t_0}^{t} \langle \text{AGSP} \rangle dt}$
&\multicolumn{1}{c}{0.07} &\multicolumn{1}{c}{-0.11} &\multicolumn{1}{c}{-0.13} &\multicolumn{1}{c}{-0.64} &\multicolumn{1}{c}{-0.54} &\multicolumn{1}{c}{-0.55} \T\B \\[10pt]
&$\dfrac{\int_{t_0}^{t} \langle \text{CP} \rangle dt}{\int_{t_0}^{t} \langle \text{Residual} \rangle dt}$ 
&\multicolumn{1}{c}{-0.33} &\multicolumn{1}{c}{-0.5} &\multicolumn{1}{c}{-0.8} &\multicolumn{1}{c}{-3.2} &\multicolumn{1}{c}{-3} &\multicolumn{1}{c}{-4.43} \\
\midrule
\multirow{4}{*}{No} 
&$\dfrac{\int_{t_0}^{t} \langle \text{CP} \rangle dt}{\int_{t_0}^{t} \langle \text{DSP} \rangle dt}$
&\multicolumn{1}{c}{0.19} &\multicolumn{1}{c}{0.25} &\multicolumn{1}{c}{0.35} &\multicolumn{1}{c}{1.46} &\multicolumn{1}{c}{1.42} &\multicolumn{1}{c}{1.68} \T \\[10pt]
&$\dfrac{\int_{t_0}^{t} \langle \text{CP} \rangle dt}{\int_{t_0}^{t} \langle \text{AGSP} \rangle dt}$
&\multicolumn{1}{c}{-0.08} &\multicolumn{1}{c}{-0.11} &\multicolumn{1}{c}{-0.14} &\multicolumn{1}{c}{-0.66} &\multicolumn{1}{c}{-0.57} &\multicolumn{1}{c}{-0.58} \T \\[10pt]
&$\dfrac{\int_{t_0}^{t} \langle \text{CP} \rangle dt}{\int_{t_0}^{t} \langle \text{Residual} \rangle dt}$ 
&\multicolumn{1}{c}{-0.35} &\multicolumn{1}{c}{-0.53} &\multicolumn{1}{c}{-0.78} &\multicolumn{1}{c}{-3.25} &\multicolumn{1}{c}{-3.06} &\multicolumn{1}{c}{-4.51} \\
\bottomrule
\end{tabular}
\end{table}

\bibliographystyle{jfm}
\bibliography{bibfile}

\begin{thebibliography}{34}
\expandafter\ifx\csname natexlab\endcsname\relax\def\natexlab#1{#1}\fi
\def\au#1{#1} \def\ed#1{#1} \def\yr#1{#1}\def\at#1{#1}\def\jt#1{\textit{#1}}
  \def\bt#1{#1}\def\bvol#1{\textbf{#1}} \def\vol#1{#1} \def\pg#1{#1}
  \def\publ#1{#1}\def\arxiv#1{#1}\def\org#1{#1}\def\st#1{\textit{#1}}

\bibitem[Asselin \& Young(2019)]{asselin2019improved}
{\sc \au{Asselin, Olivier} \& \au{Young, William~R}} \yr{2019}  \at{An improved
  model of near-inertial wave dynamics}.  \jt{Journal of Fluid Mechanics}
  \bvol{876}.

\bibitem[Barkan {\em et~al.\/}(2019)Barkan, Molemaker, Srinivasan, McWilliams
  \& D'Asaro]{barkan2019role}
{\sc \au{Barkan, Roy}, \au{Molemaker, M~Jeroen}, \au{Srinivasan, Kaushik},
  \au{McWilliams, James~C} \& \au{D'Asaro, Eric~A}} \yr{2019}  \at{The role of
  horizontal divergence in submesoscale frontogenesis}.  \jt{Journal of
  Physical Oceanography}  \bvol{49}~(6),  \pg{1593--1618}.

\bibitem[Barkan {\em et~al.\/}(2021)Barkan, Srinivasan, Yang, McWilliams, Gula
  \& Vic]{barkan2021oceanic}
{\sc \au{Barkan, Roy}, \au{Srinivasan, Kaushik}, \au{Yang, Luwei},
  \au{McWilliams, James~C}, \au{Gula, Jonathan} \& \au{Vic, Cl{\'e}ment}}
  \yr{2021}  \at{Oceanic mesoscale eddy depletion catalyzed by internal waves}.
   \jt{Geophysical Research Letters}  \bvol{48}~(18),  \pg{e2021GL094376}.

\bibitem[Barkan {\em et~al.\/}(2017)Barkan, Winters \&
  McWilliams]{barkan2017stimulated}
{\sc \au{Barkan, Roy}, \au{Winters, Kraig~B} \& \au{McWilliams, James~C}}
  \yr{2017}  \at{Stimulated imbalance and the enhancement of eddy kinetic
  energy dissipation by internal waves}.  \jt{Journal of Physical Oceanography}
   \bvol{47}~(1),  \pg{181--198}.

\bibitem[Blumen(2000)]{blumen2000inertial}
{\sc \au{Blumen, William}} \yr{2000}  \at{Inertial oscillations and
  frontogenesis in a zero potential vorticity model}.  \jt{Journal of physical
  oceanography}  \bvol{30}~(1),  \pg{31--39}.

\bibitem[Burns {\em et~al.\/}(2020)Burns, Vasil, Oishi, Lecoanet \&
  Brown]{burns2020dedalus}
{\sc \au{Burns, Keaton~J}, \au{Vasil, Geoffrey~M}, \au{Oishi, Jeffrey~S},
  \au{Lecoanet, Daniel} \& \au{Brown, Benjamin~P}} \yr{2020}  \at{Dedalus: A
  flexible framework for numerical simulations with spectral methods}.
  \jt{Physical Review Research}  \bvol{2}~(2),  \pg{023068}.

\bibitem[Chen {\em et~al.\/}(2021)Chen, Straub \& Nadeau]{chen2021interaction}
{\sc \au{Chen, Yanxu}, \au{Straub, David} \& \au{Nadeau, Louis-Philippe}}
  \yr{2021}  \at{Interaction of nonlinear ekman pumping, near-inertial
  oscillations, and geostrophic turbulence in an idealized coupled model}.
  \jt{Journal of Physical Oceanography}  \bvol{51}~(3),  \pg{975--987}.

\bibitem[Ferrari \& Wunsch(2009)]{ferrari2009ocean}
{\sc \au{Ferrari, Raffaele} \& \au{Wunsch, Carl}} \yr{2009}  \at{Ocean
  circulation kinetic energy: Reservoirs, sources, and sinks}.  \jt{Annual
  Review of Fluid Mechanics}  \bvol{41}.

\bibitem[Gerkema \& Shrira(2005)]{gerkema2005near}
{\sc \au{Gerkema, Theo} \& \au{Shrira, Victor~I}} \yr{2005}  \at{Near-inertial
  waves in the ocean: beyond the'traditional approximation'}.  \jt{Journal of
  Fluid Mechanics}  \bvol{529},  \pg{195}.

\bibitem[Gertz \& Straub(2009)]{gertz2009near}
{\sc \au{Gertz, Aaron} \& \au{Straub, David~N}} \yr{2009}  \at{Near-inertial
  oscillations and the damping of midlatitude gyres: A modeling study}.
  \jt{Journal of physical oceanography}  \bvol{39}~(9),  \pg{2338--2350}.

\bibitem[Gula {\em et~al.\/}(2014)Gula, Molemaker \&
  McWilliams]{gula2014submesoscale}
{\sc \au{Gula, Jonathan}, \au{Molemaker, M~Jeroen} \& \au{McWilliams, James~C}}
  \yr{2014}  \at{Submesoscale cold filaments in the gulf stream}.  \jt{Journal
  of Physical Oceanography}  \bvol{44}~(10),  \pg{2617--2643}.

\bibitem[Hoskins(1982)]{hoskins1982mathematical}
{\sc \au{Hoskins, Brian~J}} \yr{1982}  \at{The mathematical theory of
  frontogenesis}.  \jt{Annual review of fluid mechanics}  \bvol{14}~(1),
  \pg{131--151}.

\bibitem[Hoskins \& Bretherton(1972)]{hoskins1972atmospheric}
{\sc \au{Hoskins, Brian~J} \& \au{Bretherton, Francis~P}} \yr{1972}
  \at{Atmospheric frontogenesis models: Mathematical formulation and solution}.
   \jt{Journal of the atmospheric sciences}  \bvol{29}~(1),  \pg{11--37}.

\bibitem[Jing {\em et~al.\/}(2017)Jing, Wu \& Ma]{jing2017energy}
{\sc \au{Jing, Zhao}, \au{Wu, Lixin} \& \au{Ma, Xiaohui}} \yr{2017}  \at{Energy
  exchange between the mesoscale oceanic eddies and wind-forced near-inertial
  oscillations}.  \jt{Journal of Physical Oceanography}  \bvol{47}~(3),
  \pg{721--733}.

\bibitem[McWilliams(2016)]{mcwilliams2016submesoscale}
{\sc \au{McWilliams, James~C}} \yr{2016}  \at{Submesoscale currents in the
  ocean}.  \jt{Proceedings of the Royal Society A: Mathematical, Physical and
  Engineering Sciences}  \bvol{472}~(2189),  \pg{20160117}.

\bibitem[M{\"u}ller {\em et~al.\/}(2005)M{\"u}ller, McWilliams \&
  Molemaker]{MJJ05}
{\sc \au{M{\"u}ller, P.}, \au{McWilliams, J.~C.} \& \au{Molemaker, M.~J.}}
  \yr{2005}  \at{Routes to dissipation in the ocean: The 2d/3d turbulence
  conundrum}.  \bt{In {\em Marine Turbulence\/} (ed. \ed{H.Z. Baumert~J.
  Simpson \& J.~S{\"u}ndermann})},  \pg{pp. 397--405}.  \publ{Cambridge
  University Press}.

\bibitem[Rocha {\em et~al.\/}(2018)Rocha, Wagner \& Young]{rocha2018stimulated}
{\sc \au{Rocha, Cesar~B}, \au{Wagner, Gregory~L} \& \au{Young, William~R}}
  \yr{2018}  \at{Stimulated generation: Extraction of energy from balanced flow
  by near-inertial waves}.  \jt{Journal of Fluid Mechanics}  \bvol{847}.

\bibitem[Salmon(1980)]{salmon1980baroclinic}
{\sc \au{Salmon, Rick}} \yr{1980}  \at{Baroclinic instability and geostrophic
  turbulence}.  \jt{Geophysical \& Astrophysical Fluid Dynamics}
  \bvol{15}~(1),  \pg{167--211}.

\bibitem[Shakespeare \& Taylor(2013)]{shakespeare2013generalized}
{\sc \au{Shakespeare, Callum~J} \& \au{Taylor, John~R}} \yr{2013}  \at{A
  generalized mathematical model of geostrophic adjustment and frontogenesis:
  uniform potential vorticity}.  \jt{Journal of fluid mechanics}  \bvol{736},
  \pg{366}.

\bibitem[Srinivasan {\em et~al.\/}(2021)Srinivasan, Barkan \&
  McWilliams]{srinivasan2021cascade}
{\sc \au{Srinivasan, Kaushik}, \au{Barkan, Roy} \& \au{McWilliams, James~C}}
  \yr{2021}  \at{A forward energy cascade at fronts driven by horizontal strain
  and convergence}.  \jt{submitted} .

\bibitem[Taylor \& Straub(2016)]{taylor2016forced}
{\sc \au{Taylor, Stephanne} \& \au{Straub, David}} \yr{2016}  \at{Forced
  near-inertial motion and dissipation of low-frequency kinetic energy in a
  wind-driven channel flow}.  \jt{Journal of Physical Oceanography}
  \bvol{46}~(1),  \pg{79--93}.

\bibitem[Thomas \& Arun(2020)]{thomas2020near}
{\sc \au{Thomas, Jim} \& \au{Arun, S}} \yr{2020}  \at{Near-inertial waves and
  geostrophic turbulence}.  \jt{Physical Review Fluids}  \bvol{5}~(1),
  \pg{014801}.

\bibitem[Thomas \& Daniel(2021)]{thomas2021forward}
{\sc \au{Thomas, Jim} \& \au{Daniel, Don}} \yr{2021}  \at{Forward flux and
  enhanced dissipation of geostrophic balanced energy}.  \jt{Journal of Fluid
  Mechanics}  \bvol{911}.

\bibitem[Thomas(2012)]{thomas2012effects}
{\sc \au{Thomas, Leif~N}} \yr{2012}  \at{On the effects of frontogenetic strain
  on symmetric instability and inertia--gravity waves}.  \jt{Journal of Fluid
  Mechanics}  \bvol{711},  \pg{620--640}.

\bibitem[Thomas(2019)]{thomas2019enhanced}
{\sc \au{Thomas, Leif~N}} \yr{2019}  \at{Enhanced radiation of near-inertial
  energy by frontal vertical circulations}.  \jt{Journal of Physical
  Oceanography}  \bvol{49}~(9),  \pg{2407--2421}.

\bibitem[Thomas {\em et~al.\/}(2008)Thomas, Tandon \& Mahadevan]{TTM08}
{\sc \au{Thomas, L.~N.}, \au{Tandon, A.} \& \au{Mahadevan, A.}} \yr{2008}
  \at{Submesoscale processes and dynamics}.  \bt{In {\em Ocean {M}odeling in
  and {E}ddying {R}egime\/} (ed. \ed{M.~Hecht \& H.~Hasumi})}, ,  \vol{vol.
  177},  \pg{pp. 17--38}.  \publ{AGU Geophysical Monograph Series}.

\bibitem[Wagner \& Young(2016)]{wagner2016three}
{\sc \au{Wagner, GL} \& \au{Young, WR}} \yr{2016}  \at{A three-component model
  for the coupled evolution of near-inertial waves, quasi-geostrophic flow and
  the near-inertial second harmonic}.  \jt{Journal of Fluid Mechanics}
  \bvol{802},  \pg{806}.

\bibitem[Weller(1982)]{weller1982relation}
{\sc \au{Weller, Robert~A}} \yr{1982}  \at{The relation of near-inertial
  motions observed in the mixed layer during the jasin (1978) experiment to the
  local wind stress and to the quasi-geostrophic flow field}.  \jt{Journal of
  Physical Oceanography}  \bvol{12}~(10),  \pg{1122--1136}.

\bibitem[Whitt \& Thomas(2013)]{whitt2013near}
{\sc \au{Whitt, Daniel~B} \& \au{Thomas, Leif~N}} \yr{2013}  \at{Near-inertial
  waves in strongly baroclinic currents}.  \jt{Journal of physical
  oceanography}  \bvol{43}~(4),  \pg{706--725}.

\bibitem[Whitt \& Thomas(2015)]{whitt2015resonant}
{\sc \au{Whitt, Daniel~B} \& \au{Thomas, Leif~N}} \yr{2015}  \at{Resonant
  generation and energetics of wind-forced near-inertial motions in a
  geostrophic flow}.  \jt{Journal of Physical Oceanography}  \bvol{45}~(1),
  \pg{181--208}.

\bibitem[Winters \& de~la Fuente(2012)]{winters2012modelling}
{\sc \au{Winters, Kraig~B} \& \au{de~la Fuente, Alberto}} \yr{2012}
  \at{Modelling rotating stratified flows at laboratory-scale using
  spectrally-based dns}.  \jt{Ocean Modelling}  \bvol{49},  \pg{47--59}.

\bibitem[Xie(2020)]{xie2020downscale}
{\sc \au{Xie, Jin-Han}} \yr{2020}  \at{Downscale transfer of quasigeostrophic
  energy catalyzed by near-inertial waves}.  \jt{Journal of Fluid Mechanics}
  \bvol{904}.

\bibitem[Xie \& Vanneste(2015)]{xie2015generalised}
{\sc \au{Xie, J-H} \& \au{Vanneste, Jacques}} \yr{2015}  \at{A
  generalised-lagrangian-mean model of the interactions between near-inertial
  waves and mean flow}.  \jt{Journal of Fluid Mechanics}  \bvol{774},
  \pg{143--169}.

\bibitem[Young \& Jelloul(1997)]{young1997propagation}
{\sc \au{Young, WR} \& \au{Jelloul, Mahdi~Ben}} \yr{1997}  \at{Propagation of
  near-inertial oscillations through a geostrophic flow}.  \jt{Journal of
  marine research}  \bvol{55}~(4),  \pg{735--766}.

\end{thebibliography}

\end{document}